\documentclass[12pt]{article}
\usepackage{cite}
\usepackage{amsbsy}
              
\title{Lagrangian and Presymplectic Particle Dynamics with Maximal Acceleration.}
\author{M. Toller \thanks{e-mail: toller@iol.it}\\ 
via Malfatti n. 8  \\
I-38100 Trento, Italy}
 
\begin{document} 
\maketitle
                 
\begin{abstract}
We discuss some Lagrangian and presymplectic models concerning test particles in electromagnetic and gravitational fields, with the aim of describing an upper bound to the acceleration. Some models are based on the relativistic phase space and others on the bundle of the Lorentz frames. For the second case, an appropriate version of the methods of analytic mechanics, including the Noether theorem, is developed. A strict application of the analogy with the bound to velocity which appears in relativity theory gives rise to interesting models which, however, have an unphysical energy-momentum spectrum or do not imply the required upper bound. With some modifications we obtain more acceptable models with a correct energy-momentum spectrum and with an upper bound to a quantity similar to the acceleration, that we call ``pseudo-acceleration''.
\bigskip

\noindent PACS numbers:

45.50.Dd (Dynamics and kinematics of a particle); 
 
45.20.Jj (Lagrangian and Hamiltonian mechanics); 

11.30.Cp (Lorentz and Poincar\'e invariance in particles and fields).  
\end{abstract}

\tableofcontents

\section{Introduction.}

In a preceding paper \cite{Toller1} we have examined some geometric structures which describe an upper bound to the acceleration of a particle measured in its rest frame. A bound of this kind has been proposed in 1981 by Caianiello \cite{Caianiello} and since then it has been discussed by several authors. Some general remarks and a long, but not complete, list of references is given in \cite{Toller1}. 

The basic idea of ref. \cite{Toller1} was a natural generalization of the usual relativistic treatment of a test particle. The motion of the particle is described by a trajectory $\tau \to s(\tau)$ in a suitable manifold $\mathcal{S}$; $\tau$ is an arbitrary parameter and $s(\tau) \in \mathcal{S}$. We indicate by
\begin{equation}
X(\tau) = \frac{d s(\tau)}{d \tau} \in \mathcal{T}(s) = T_s\mathcal{S}
\end{equation}
the vectors tangent to the trajectory, which lie in the tangent spaces of the manifold at the points $s(\tau)$. In every tangent space $\mathcal{T}(s)$ a cone $\mathcal{T}^+(s)$ is given, and we require that $X(\tau) \in \mathcal{T}^+(s(\tau))$. 

By {\it cone} we mean what other authors call a {\it convex cone}. The convexity assumption is essential in the whole treatment and, together with some natural topological and symmetry assumptions, determines the cones up to the choice of the maximal acceleration parameter $a_M = \lambda^{-1}$. Even if the maximal acceleration has probably to be considered as a quantum effect, our treatment is classical. We use a system of units in which $c = 1$.

In ref.\ \cite{Toller1} we have considered three different schemes: in two of them the manifold $\mathcal{S}$ is the eight-dimensional relativistic phase space or the tangent bundle of the space-time manifold, while in the third scheme $\mathcal{S}$ is the ten-dimensional bundle of the Lorentz frames of space-time.

In the present paper, we discuss the dynamical aspects of the problem, namely the changes that have to be introduced into the equations of motion of a test particle in order to avoid the violation of the maximal acceleration principle. We shall concentrate our attention on the Lagrangian or Hamiltonian formalisms, necessary for the construction of quantum models. In the more complex models we adopt the equivalent presymplectic formalism \cite{Souriau,Kunzle}, which permits a more flexible treatment of the constraints and of the gauge symmetries \cite{Dirac}. For instance, we do not discuss an interesting model based on the bundle of the pseudo-complex Lorentz frames \cite{Schuller1,SP}, because it is not formulated with the methods of analytic mechanics.

We follow again the analogy with the relativity theory. The first step, concerns the dependence of the rate of an ideal clock on its acceleration (and possibly, since a clock is an extended object, on its angular velocity). It is clear that all the real clocks are influenced by the inertial forces due to acceleration and by the centrifugal forces due to rotation. These forces can even destroy the clock mechanism. We are dealing with a dependence that cannot be made arbitrarily small by building more and more robust mechanisms and cannot be explained in terms of the standard theory.

We assume that  the {\it proper time} $\sigma$ measured by the clock has the form
\begin{equation}
d \sigma = - m^{-1}L(X(\tau)) d \tau, 
\end{equation}
where $L(X)$ is an homogeneous function of degree one, so that the result does not depend on the choice of the parameter $\tau$. The constant factor $- m^{-1}$ has been introduced for future convenience. 

For instance in special relativity $\tau \to x(\tau)$ is a curve in the space-time manifold and we have 
\begin{equation}
L = - m (\dot x \cdot \dot x)^{1/2},
\end{equation}
where $x \cdot y = x_k y^k = x^0 y^0 - \mathbf{x} \cdot \mathbf{y}$ is the relativistic scalar product of two four-vectors and the dot indicates the derivative with respect to $\tau$. The same formula holds in the presence of gravitation, but then the scalar product is defined in terms of the metric tensor.

Guided by this formula, we require that the function $L(X)$ vanishes when $X$ approaches the boundary of the cone $\mathcal{T}^+(s)$ and has the same symmetry group as the cone. In some cases, as in the usual relativity theory, these properties and the homogeneity requirement determine the form of $L(X)$ up to a constant multiplicative factor.

The second step, always suggested by relativity, is to identify the function $L(X)$ with the Lagrangian which appears in the action principle for a free test particle
\begin{equation} \label{Action}
\delta \int_{\tau_1}^{\tau_2} L(X(\tau)) \, d \tau = 0
\end{equation}
and to study the Euler-Lagrange equations.  

In order to have a nonvanishing acceleration, we have to introduce some classical external fields and to add to the free Lagrangian $L$ a suitable interaction Lagrangian $L_I$. In the present paper we consider only gravitational and electromagnetic fields described by Einstein's and Maxwell's theories. We remark, however, that a more general approach to the test particle dynamics may open the way to the treatment of a wider class of long-rage classical fields. A list of the new long-range fields proposed by many authors would be too long or incomplete and we shall treat this argument elsewhere.

In the following we apply the ideas explained above to the geometries introduced in ref.\ \cite{Toller1}. In section 2 we give a short discussion of a model based on the relativistic phase space, which has been treated with more detail and with a different method in ref.\ \cite{NFS}. It is known that this model has an unphysical energy-momentum spectrum, but we discuss it because it suggests some important concepts useful for the construction of more acceptable models.

The section 3 is dedicated to some formal developments of the methods of Lagrangian mechanics applied to the bundle $\mathcal{S}$ of the Lorentz frames. In Section 4, following the indications of ref.\ \cite{Kunzle}, we present, in the same geometric framework, a general presymplctic formalism. Both treatments include a discussion of the symmetries and of the conservation laws (Noether theorem). In section 5 we discuss the relation between a trajectory in $\mathcal{S}$ and the world line of the particle in the spacetime manifold $\mathcal{M}$.

In section 6 we use the formalism developed in the preceding sections to introduce, in the bundle of Lorentz frames, a model based on a direct analogy with relativity theory. It has a physically acceptable energy-momentum spectrum, but the particle acceleration is not bounded.

The models discussed in sections 2 and 6 do not describe elementary systems (with the Poincar\'e group acting transitively on the phase space), but systems with additional ``internal'' degrees of freedom. Another feature of both  models is that the ``distinguished point'' of the system, which appears in the equations  of motion and to which the external forces are applied, does not coincide, in general, with the center of mass. This fact is another indication of the extended nature of the particle and it raises the issue whether the maximal acceleration principle refers to the ``distinguished point'', as in the model of section 2, or, for example, to the center of mass.

In section 7 we exploit the experience gained in the preceding sections for the construction of new models of two kinds.  The models of the first kind provide an upper bound to the acceleration, but admit spacelike energy-momentum four-vectors. In particular we discuss again the model described in section 2. The models of the second kind have a physically acceptable energy-momentum spectrum, but the upper bound does not concern the acceleration, but a different quantity called the ``pseudo-acceleration'' introduced in section 5. The existence of physically acceptable Lagrangian or presymplecxtic models with an upper bound to the ``true'' acceleration remains an open problem.

We also consider models for spinning particles which, besides a maximal acceleration or pseudo-acceleration, imply an upper bound to the angular velocity of the spin precession. We do not find in these models the correction to the Thomas precession proposed, on the basis of a different model, in ref.\ \cite{Schuller1}. 

It is clear that many other similar models can be proposed. In the present investigation the maximal acceleration (or pseudo-acceleration) seems to appear as a property of some kinds of extended particles rather than an intrinsic fundamental feature of the spacetime structure. Perhaps, this conclusion is inavoidable in a scheme which disregards quantum effects and considers gravitation as an external field.

\section{Dynamics in the relativistic phase space.}

\subsection{The choice of the Lagrangian.}

In the present section we summarize the first two kinematical schemes considered in ref.\ \cite{Toller1}, we consider explicit formulas for the rate of accelerated ideal clocks and we study the action principle and the dynamical equations which describe the motion of a test particle. These schemes are both based on the tangent bundle $T \mathcal{M}$ of the spacetime $\mathcal{M}$. Since $\mathcal{M}$ has a pseudo-Riemannian metric, $T \mathcal{M}$ can be identified with the cotangent bundle $T^* \mathcal{M}$, namely with the relativistic phase space.

Detailed treatments in the presence of gravitational and electromagnetic fields can be found in the literature \cite{NFS,Scarpetta,Brandt,NFS2,FLPS,Schuller}. There are several different approaches, but we concentrate the attention on a specific Lagrangian model which follows directly from the ideas presented in the introduction.

In the present section we consider a flat Minkowski spacetime. In the presence of a gravitational field (or if we want simply to use general spacetime coordinates), we have to modify the formulas by introducing in the appropriate way the metric tensor and the connection coefficients. Another approach is summarized in section 7.5.

We indicate by $x^k$ the coordinates and by
\begin{equation} \label{DefU}
u = (\dot x \cdot \dot x)^{-1/2} \dot x
\end{equation}
the four-velocity. If $\tau$ is the usual proper time of relativity (different from the new proper time $\sigma$), we simply have $u = \dot x$.  In any case we have the constraint
\begin{equation} \label{Shell}
u \cdot u = 1.
\end{equation}

We introduce in the eight-dimensional vector space $T \mathcal{M}$ the coordinates $(x, u)$ and we indicate by $\mathcal{V}$ the seven-dimensional submanifold defined by the constraint (\ref{Shell}), called the space-time-velocity space.  The motion of a particle is described by a trajectory in the space $\mathcal{V}$ or, if one prefers, in $T \mathcal{M}$. We alway choose the parameter $\tau$ in such a way that $u^0 > 0$.

In the first geometric scheme, which is based on Born's reciprocity principle \cite{Born1,Born2} and has been developed in refs.\ \cite{Caianiello,CDFMV,Brandt1} and in many other papers, the maximal acceleration is described by the inequality
\begin{equation} \label{Cond1}
dx \cdot dx + \lambda^2 du \cdot du \geq 0.
\end{equation}
It does not define a cone in $T \mathcal{M}$, but it defines cones in the tangent spaces of $\mathcal{V}$. From the motivations explained in the introduction, we get immediately
\begin{equation} \label{Lagr1}
L(\dot x, \dot u) = - m (\dot x \cdot \dot x + \lambda^2 \dot u \cdot \dot u)^{1/2}.
\end{equation}

The second geometric scheme, inspired by the Born-Infeld theory of electromagnetism \cite{BI}, has recently been introduced in ref.\ \cite{Schuller}.  The maximal acceleration is described by the inequalities
\begin{equation} \label{Cond2}
dx_+ = dx + \lambda \, du \in V^+, \qquad dx_- = dx - \lambda \, du \in V^+,
\end{equation}
where $V^+$ is the usual closed future cone of special relativity.  It is clear that this formula defines cones both in $T \mathcal{M}$ and in the tangent spaces of $\mathcal{V}$. It has been shown in ref.\ \cite{Toller1} that the cone in $T \mathcal{M}$ is uniquely determined, up to the choice of the constant $a_M = \lambda^{-1}$, by its Lorentz symmetry and some other simple and natural properties. The requirements given in the introduction are satisfied by the formula
\begin{eqnarray} \label{Lagr2}
& L(\dot x, \dot u) = - m (\dot x_+ \cdot \dot x_+)^{1/4} (\dot x_- \cdot \dot x_-)^{1/4} = & \nonumber  \\
&- m \left((\dot x \cdot \dot x + \lambda^2 \dot u \cdot \dot u)^2 - 4\lambda^2 (\dot x \cdot \dot u)^2 \right)^{1/4}.&
\end{eqnarray}

If we take into account the definition (\ref{DefU}), we have
\begin{equation} 
\dot x \cdot \dot u = 0
\end{equation}
and both the Lagrangians (\ref{Lagr1})and (\ref{Lagr2}) can be written in the form
\begin{equation} \label{Lagr3}
L(\dot x, \ddot x) = - m \left(\dot x \cdot \dot x + \lambda^2 (\dot x \cdot \dot x)^{-1} \ddot x \cdot \ddot x - \lambda^2 (\dot x \cdot \dot x)^{-2} (\ddot x \cdot \dot x)^2 \right)^{1/2}.
\end{equation}
The same Lagrangian follows from an approach based on Clifford algebras \cite{Castro}. 

\subsection{The dynamical equations.}

The Lagrange equations following from the action principle (\ref{Action}) have the form
\begin{equation} \label{Dyn1}
\frac{d^2}{d \tau^2} \frac{\partial L}{\partial \ddot x^k}
- \frac{d}{d \tau} \frac{\partial L}{\partial \dot x^k} = 0.
\end{equation}
This is a system of fourth order differential equations invariant under changes of the parameter $\tau$ and under the Poincar\'e group. 

From the Noether theorem we find ten conserved quantities \cite{NFS,CGZ}, namely the energy-momentum four-vector
\begin{equation} \label{FMom}
p_k = - \dot\pi_k - \frac{\partial L}{\partial \dot x^k}
\end{equation}
and the relativistic angular momentum
\begin{equation} 
m_{ik} = x_i p_k - x_k p_i + \dot x_i \pi_k - \dot x_k \pi_i,
\end{equation}
where
\begin{equation} 
\pi_k = - \frac{\partial L}{\partial \ddot x^k} =
- m^2 \lambda^2 L^{-1} (\dot x \cdot \dot x)^{-1}
\left(\ddot x_k -  (\dot x \cdot \dot x)^{-1} (\ddot x \cdot \dot x) \dot x_k\right).
\end{equation}
The non conserved quantities
\begin{equation} 
p_{ik} = \dot x_i \pi_k - \dot x_k \pi_i = 
- m^2 \lambda^2 L^{-1} (\dot x \cdot \dot x)^{-1}
\left(\dot x_i\ddot x_k - \dot x_k\ddot x_i\right)
\end{equation}
represent the relativistic angular momentum measured with respect to the point $x(\tau)$. 

The space coordinates of the center of mass at the time $x^0$ are given by \cite{Moller}
\begin{equation} \label{COM}
y_r = p_0^{-1} m_{r0} + x^0 p_0^{-1} p_r = x_r + p_0^{-1} p_{r0}, \qquad r = 1, 2, 3.
\end{equation}
Note that the world line of the center of mass may depend on the Lorentz frame. The velocity of the center of mass is given by
\begin{equation} \label{VelCOM}
\mathbf{v} = p_0^{-1} \mathbf{p}.
\end{equation}
The Pauli-Lubanski four-vector, which describes the spin of the particle, is given by
\begin{equation} \label{PL}
S^k = 2^{-1} \epsilon^{kijl}m_{ij} p_l =
2^{-1} \epsilon^{kijl}p_{ij} p_l =
\epsilon^{kijl} \dot x_i \pi_j p_l.
\end{equation}
We see that the center of mass may not coincide with the ``distinguished point'' $x(\tau)$ and the particle may have a nonvanishing spin \cite{NFS}. 

A detailed treatment of the equations of motion and explicit general solutions are given in ref.\ \cite{NFS} (see also section 7.5). Here we consider some simple aspects which are useful for the physical interpretation.  We simplify the calculations by choosing the parameter $\tau = x^0$. Of course, the Lorentz symmetry becomes less evident. We have $\dot x^0 = 1$ and the Lagrangian takes the form
\begin{eqnarray}
&L(\dot \mathbf{x}, \ddot\mathbf{x}) = - m & 
\nonumber \\
&\times \left(1 - \|\dot\mathbf{x}\|^2 - \lambda^2 (1 - \|\dot\mathbf{x}\|^2)^{-1} \|\ddot\mathbf{x}\|^2 - \lambda^2 (1 - \|\dot\mathbf{x}\|^2)^{-2} (\ddot\mathbf{x} \cdot \dot\mathbf{x})^2 \right)^{1/2}.&
\end{eqnarray}
We obtain three fourth order differential equations for the components of the vector $\mathbf x$ similar to eq.\ (\ref{Dyn1}). We always assume that $\dot\mathbf x$ and $\ddot{\mathbf x}$ satisfy the upper bounds to velocity and acceleration, so that the Lagrangian is real. 

After some calculations we find that the $3 \times 3$ matrix
\begin{equation} \label{Matrix}
A_{rs} = \frac{\partial^2 L}{\partial \ddot x^r \ddot x^s} 
= \frac{\partial}{\partial \stackrel{....}{x}^r} \frac{d^2}{d \tau^2}
\frac{\partial L}{\partial \ddot x^s},
\qquad r, s = 1, 2, 3 
\end{equation}
is definite positive and therefore nonsingular. It follows that the dynamical equations determine the fourth derivatives $\stackrel{....}{\mathbf {x}}$ uniquely as functions of the lower order derivatives. In other words, the differential equations can be written in normal form and the uniqueness and existence theorems can be applied. It follows that, at least locally, the world line is determined by the initial conditions $\mathbf{x}(0)$, $\dot\mathbf{x}(0)$, $\ddot\mathbf{x}(0)$ and $\stackrel{...}{\mathbf{x}}(0)$. The phase space of the dynamical system has dimension twelve, but the homogeneous symplectic spaces of the Poincar\'e group \cite{Souriau} have at most dimension eight. This means that the Poincar\'e group does not act transitively on the phase space and the model does not describe an ``elementary'' particle.

It is easy to see that a linear timelike world line, namely a line that, with an appropriate choice of the parameter, is characterized by a constant timelike value of $\dot x(\tau)$, is a solution of the equations of motion. In this case we have
\begin{equation} 
\pi_k = 0, \qquad p_k = m (\dot x \cdot \dot x)^{-1/2} \dot x_k, \qquad
p_{ik} = 0, \qquad S_k = 0.
\end{equation}
These solutions describe a spinless particle with mass $m$ and with the center of mass coinciding with $x(\tau)$. A similar result holds in the presence of a gravitational field, as it is shown in section 7.5. In other words, if there are only gravitational fields, the world lines calculated by means of Einstein's theory are also solutions of the present model and there is no correction due to the maximal acceleration. For a different point of view, see ref.\ \cite{CFLPA}.

In a solution of the general kind, the four-velocity $\dot x$ is a timelike four-vector and the dynamical quantities at a given value of the parameter, for instance $\tau = x^0 = 0$, can be further simplified by working in the rest frame in which we have $\dot\mathbf{x}(0) = 0$. Remember that with our choice of the parameter $\tau$ we have $\dot x^0 = 1$ and all the higher derivatives of $x^0(\tau)$ vanish. With some calculations we obtain for $\tau = 0$
\begin{eqnarray} \label{Rest}
&L = -m (1 - \lambda^2 \, \|\ddot{\mathbf{x}}\|^2)^{1/2}, \qquad
p_0 = m (1 - \lambda^2 \, \|\ddot{\mathbf{x}}\|^2)^{-1/2}\geq m,& 
\nonumber \\
&\mathbf{p} = - \lambda^2 p_0 \stackrel{...}{\mathbf{x}} 
- \lambda^4 m^{-2} p_0^3 \, (\ddot{\mathbf{x}} \cdot \stackrel{...}{\mathbf{x}}) \ddot{\mathbf{x}},&
\nonumber \\
&p_{rs} = 0, \qquad p_{r0} =  \lambda^2 p_0 \ddot x_r, \qquad r, s, = 1, 2, 3.&
\end{eqnarray}
By an appropriate choice of $\ddot{\mathbf{x}}$ and $\stackrel{...}{\mathbf{x}}$ we can obtain any value of $p^0 \geq m$ and of $\|\mathbf{p}\|$. The other points of the energy-momentum spectrum can be obtained by means of a Lorentz transformation.  It follows that the spectrum is the complement of the closed past cone and contains all the (unphysical) spacelike four-momenta. All the real values of the invariant $(p \cdot p)$ are admitted.

From eq.\ (\ref{COM}) we obtain the following simple expression, valid in a rest frame, for the coordinates of the center of mass 
\begin{equation} 
\mathbf{y} = \mathbf{x} - \lambda^2 \ddot \mathbf{x}.
\end{equation}
Since, in order to have a real finite energy $p_0$, we must have $\lambda \|\ddot \mathbf{x}\| < 1$, we obtain the upper bound, 
\begin{equation} 
\|\mathbf{y} - \mathbf{x}\| <  \lambda.
\end{equation}

In conclusion, we have seen that a linear world line describes a relativistic spinless particle with mass $m$, but there are other solutions with different mass and spin and also solutions with an unphysical spacelike energy-momentum.

In order to investigate the effects of acceleration, we have to introduce non-gravitational forces, for instance electromagnetic forces, described by adding to the Lagrangian (\ref{Lagr3}) the interaction Lagrangian   
\begin{equation} 
L_I = - e \dot x^k A_k(x).
\end{equation}
We are assuming that the electric charge $e$ is concentrated on the world line $x(\tau)$ and not necessarily in the center of mass. This is not the only possible kind of electromagnetic interaction and other choices, suggested by the Born-Infeld electrodynamics, are discussed in ref. \cite{Schuller}. 

With our simple choice of the interaction Lagrangian, the equations of motions take the familiar form
\begin{equation} \label{DynF}
\dot p_k = e F_{ki}(x) \dot x^i, \qquad F_{ki} = \partial_k A_i - \partial_i A_k,
\end{equation}
where, however, $p$ is given by eq.\ (\ref{FMom}). A perturbative treatment, starting from a linear solution, is described in ref.\ \cite{FLPS} for small values of the field $F_{ki}$. Here we study the solutions for small values of $\tau$, given the initial conditions $x(0)$, $\dot x(0)$, $\ddot x(0) = 0$ and $\stackrel{...}{x}(0) = 0$. In other words, we consider a free particle moving on a linear world line and we study what happens if, starting from $\tau = 0$, it interacts with an electromagnetic field. Without any loss of generality, we choose the parameter $\tau = x^0$ and we work in the rest system of the particle at $\tau = 0$. 

From eq.\ (\ref{FMom}), after some calculations, we find
\begin{equation} 
p_0(0) = m, \qquad \mathbf{p}(0) = 0, \qquad  
\dot p_0(0) = 0, \qquad\dot\mathbf{p}(0) = - m \lambda^2 \stackrel{....}{\mathbf{x}}(0)
\end{equation}
and the equation of motion at $\tau = 0$ takes the form
\begin{equation} 
\stackrel{....}{\mathbf{x}}(0) = - \lambda^{-2} m^{-1} e \mathbf{E}(x(0)),
\end{equation}
where $\mathbf{E}$ is the electric field with components $F^{r0}$. Note that the point $x(\tau)$ start its motion in the wrong direction and the field does a negative work.

The velocity of the center of mass is given by eq.\ (\ref{VelCOM}) and  we obtain the classical formula
\begin{equation} 
\mathbf{v}(0) = 0, \qquad
\dot\mathbf{v}(0) = m^{-1} e \mathbf{E}(x(0)).
\end{equation}
We see that the acceleration of the center of mass has no upper bound. Since the point $x(\tau)$ has no acceleration at $\tau = 0$, if there is an external force, the world line of the point $x(\tau)$ does not coincide with the world line of the center of mass for $\tau > 0$.

Under the same conditions from eq.\ (\ref{DynF}) we have, disregarding higher powers of $\tau$,
\begin{equation} 
p_0 \approx m - (24)^{-1} \tau^4 \lambda^{-2} m^{-1} e^2 \|\mathbf{E}(x(0))\|^2, \qquad 
\mathbf{p} \approx \tau e \mathbf{E}(x(0)),
\end{equation}
\begin{equation} 
(p \cdot p) \approx m^2 - \tau^2 e^2 \|\mathbf{E}(x(0))\|^2.
\end{equation}
As a consequence of the external field, the mass of the particle decreases, at least initially, because its momentum increases while its energy decreases. One can show that, under some conditions, the particle also acquires a non vanishing spin.

We conclude that the model discussed in this section has some physically unacceptable properties, but it suggests some interesting ideas, in particular the distinction between the world line of the center of mass and the world line of the distinguished point $x(\tau)$ where the electric charge is concentrated or, more in general, where the external force is applied. 

\section{Lagrangian dynamics in the space of the Lorentz frames.}

\subsection{The bundle of frames.}

In the third kinematical scheme discussed in ref. \cite{Toller1}, for each value of the parameter $\tau$, we associate to the (possibly extended) particle a local Lorentz frame $s(\tau)$ in such a way that the distinguished point (not necessarily the center of mass) coincides with the origin of the frame (sometimes called a {\it moving frame}). We indicate by $\mathcal{S}$ the ten-dimensional bundle of the Lorentz frames of the space-time and we describe the motion of the particle by the line $\tau \to s(\tau) \in \mathcal{S}$.

Particle dynamics (without maximal acceleration) in this space has been introduced by K\"unzle \cite{Kunzle} (see also \cite{Vanzo}) in the framework of the presymplectic formalism \cite{Souriau}. A treatment based on the balance equations for the densities of energy, momentum and relativistic angular momentum, strongly influenced by Dixon's work \cite{Dixon}, is given in refs.\ \cite{SZ,Toller2,TV}. In the present section we develop a geometric Lagrangian formalism which  avoids the introduction of coordinate systems in the space $\mathcal{S}$. It will be applied in the following sections to the maximal acceleration dynamics. It is more general than the Lagrangian formalism used in ref.\ \cite {NFS}, because $s(\tau)$ is not assumed to be a rest frame. We shall see that this feature permits the construction of interesting models.

The elements of $\mathcal{S}$  are orthonormal tetrads $\{e_0,\ldots, e_3\}$ of four-vectors in the pseudo-Riemannian space-time manifold $\mathcal{M}$. We assume that $\mathcal{M}$ is time-oriented and $e_0$ belongs to the future cone. We consider ten vector fields $A_0,\ldots, A_9$ in the manifold $\mathcal{S}$.  The fields  $A_0,\ldots, A_3$ generate parallel displacements of the tetrads along the directions of the tetrad vectors, $A_4 = A_{[23]}$, $A_5 = A_{[31]}$, $A_6 = A_{[12]}$ generate rotations around the spatial vectors of the tetrad and $A_7 = A_{[10]}$, $A_8 = A_{[20]}$, $A_9 = A_{[30]}$ generate Lorentz boosts along the same spatial vectors. The latin indices $i, k, j, l, m, n$ take the values $0,\ldots, 3$ and the greek indices $\alpha, \beta, \gamma$ take the values $0,\ldots, 9$. We assume $A_{[ik]} = A_{[ki]}$ and, when necessary, we use the square brackets to indicate that an antisymmetric pair of latin indices stands for a greek index. 

We also introduce in the space $\cal S$ the differential forms $\omega^{\beta}$ dual to the vector fields $A_{\alpha}$, defined by
\begin{equation}
i_{A_{\alpha}} \omega^{\beta} = \omega^{\beta}(A_{\alpha}) = \delta_{\alpha}^{\beta},
\end{equation}
where $i_X$ is the interior product operator acting on the differential forms.

The vector fields $A_{\alpha}$ can also be considered as first order differential operators and their commutators (Lie brackets) can be written in the form
\begin{equation} \label{Coeff}
[A_{\alpha}, A_{\beta}] = F_{\alpha \beta}^{\gamma}  A_{\gamma}.
\end{equation}
The quantities $F_{\alpha \beta}^{\gamma} = - F_{\beta \alpha}^{\gamma}$, are called {\it structure coefficients} and in the absence of gravitation they are the structure constants of the Poincar\'e group. In the presence of gravitation, $F_{ik}^{[jl]}$ are the components of the curvature tensor and $F_{ik}^j$ are the components of the torsion tensor, which vanishes in Einstein's theory.

The ten dimensional manifold $\mathcal{S}$ has a structure of principal fibre bundle \cite{KN} with basis $\mathcal{M}$ and structural group $SO^{\uparrow}(1, 3)$, but the details of this structure are not needed in the next subsection. The only relevant feature is that $\mathcal{S}$ in a $n$-dimensional differentiable manifold in which $n$ differentiable vector fields $A_{\alpha}$ are defined and that, for each point $s \in \mathcal{S}$, the vectors $A_{\alpha}(s)$ form a basis of the tangent space. As a consequence, we can identify all the tangent spaces $T_s\mathcal{S}$ with a single ten-dimensional vector space $\mathcal{T}$, which, in the absence of gravitation and other external fields, is the Lie algebra of the Poincar\'e group. More details on this kind of geometry can be found in ref.\ \cite{CSTVZ}. In this way we can also treat a large class of external fields, including torsion fields \cite{HHKN}, and, for $n > 10$, gauge fields \cite{CSVZ}.

\subsection{The Lagrange equations.}

We write the tangent vectors of the curve $\tau \to s(\tau) \in \mathcal{S}$, which  describes the motion of the frame, in the form
\begin{equation} \label{Deriv}
X(\tau) = \frac{ds(\tau)}{d \tau} = b^{\alpha} A_{\alpha} 
\in \mathcal{T}
\end{equation}
and we consider the action principle 
\begin{equation}
\delta \int_{\tau_1}^{\tau_2} L(b^{\alpha}, s) \, d \tau = 0.
\end{equation} 
where the Lagrangian $L$ is an homogeneous function of degree one of the ``velocities'' $b^{\alpha}$.

In order to derive the dynamical equations, we consider a family, parametrized by $\epsilon$, of varied trajectories $(\epsilon, \tau) \to s(\epsilon, \tau)$ with the usual properties
\begin{equation} \label{Conditions}
s(0, \tau) = s(\tau), \qquad
s(\epsilon, \tau_1) = s(\tau_1), \qquad s(\epsilon, \tau_2) = s(\tau_2).
\end{equation}
We put
\begin{equation}
\frac{\partial s(\epsilon, \tau)}{\partial \epsilon} = a^{\alpha} A_{\alpha}.
\end{equation}

If $\phi(s)$ is a differentiable auxiliary function, we have
\begin{equation}
\frac{\partial \phi(s)}{\partial \epsilon} = a^{\alpha} A_{\alpha}\phi, \qquad
\frac{\partial \phi(s)}{\partial \tau} = b^{\alpha} A_{\alpha}\phi,
\end{equation}
\begin{equation}
\frac{\partial^2 \phi(s)}{\partial \epsilon \partial \tau} = 
a^{\alpha} A_{\alpha} b^{\beta} A_{\beta}\phi =
b^{\alpha} A_{\alpha} a^{\beta} A_{\beta}\phi.
\end{equation}
From the last equality we obtain
\begin{equation}
\left((a^{\alpha} A_{\alpha} b^{\beta}) A_{\beta} - 
(b^{\alpha} A_{\alpha} a^{\beta}) A_{\beta} +
a^{\alpha} b^{\beta} [A_{\alpha}, A_{\beta}] \right)\phi = 0,
\end{equation}
namely
\begin{equation}
\left(\frac{\partial b^{\gamma}}{\partial \epsilon} - 
\frac{\partial a^{\gamma}}{\partial \tau} +
a^{\alpha} b^{\beta} F_{\alpha \beta}^{\gamma} \right) A_{\gamma} \phi = 0,
\end{equation}
and, since $\phi$ is arbitrary,
\begin{equation}
\frac{\partial b^{\gamma}}{\partial \epsilon}  = 
\frac{\partial a^{\gamma}}{\partial \tau} -
a^{\alpha} b^{\beta} F_{\alpha \beta}^{\gamma}
\end{equation}
and finally (disregarding higher order terms in $\epsilon$) 
\begin{equation} \label{DeltaB}
\delta b^{\alpha} = 
\epsilon \left( \frac{\partial b^{\alpha}}{\partial \epsilon} \right)_{\epsilon = 0} = \epsilon \left( \frac{d a^{\alpha}}{d \tau} -
a^{\beta} b^{\gamma} F_{\beta \gamma}^{\alpha} \right)_{\epsilon = 0}
= \epsilon b^{\gamma} \left( A_{\gamma} a^{\alpha} -
a^{\beta} F_{\beta \gamma}^{\alpha} \right)_{\epsilon = 0}.
\end{equation}

By means of the last formula, one can write, performing a partial integration,
\begin{eqnarray} \label{Variation}
&\delta \int_{\tau_1}^{\tau_2} L \, d \tau =
\epsilon \int_{\tau_1}^{\tau_2} \left(\frac{\partial L}{\partial b^{\alpha}} 
\left( \frac{d a^{\alpha}}{d \tau} -
a^{\beta} b^{\gamma} F_{\beta \gamma}^{\alpha} \right) 
+ a^{\alpha} A_{\alpha} L \right)\, d \tau =& \nonumber \\
&\epsilon \int_{\tau_1}^{\tau_2} \left(- \frac{d}{d \tau} \frac{\partial L}{\partial b^{\alpha}} 
- \frac{\partial L}{\partial b^{\beta}}
b^{\gamma} F_{\alpha \gamma}^{\beta} +  A_{\alpha} L \right) a^{\alpha} \, d \tau + 
\epsilon \left[\frac{\partial L}{\partial b^{\alpha}} a^{\alpha} \right]_{\tau_1}^{\tau_2}.&
\end{eqnarray} 
The last term vanishes as a consequence of the conditions (\ref{Conditions}) and, considering that $a^{\alpha}$ is an arbitrary function of $\tau$, we obtain the Euler-Lagrange dynamical equations
\begin{equation} \label{Dyn3}
\dot p_{\alpha} = \frac{d p_{\alpha}}{d \tau} =
b^{\gamma} p_{\beta} F_{\gamma \alpha}^{\beta} - A_{\alpha} L,
\end{equation} 
where
\begin{equation} \label{Dyn4}
p_{\alpha} = - \frac{\partial L}{\partial b^{\alpha}}. 
\end{equation} 

If we assume that $L$ does not depend directly on $s$, the last term in eq.\ (\ref{Dyn3}) is not present and this equation is exactly the one obtained in ref.\ \cite{Toller2} in the pole approximation by integrating the balance equations (the quantities $F_{\alpha \beta}^{\gamma}$ are defined there with a different sign). 

The momenta $p_{\alpha}$ defined by eq.\ (\ref{Dyn4}) are homogeneous functions of degree zero of the velocities $b^{\alpha}$, namely they depend only on the ratios $(b^0)^{-1} b^{\alpha}$ (and possibly on $s$). It follows that they must satisfy at least a primary constraint \cite{Dirac}.    Other primary constraints may exist and we write them in the form
\begin{equation} \label{Constraints}
\Phi_{\rho}(p_{\alpha}, s) = 0, \qquad \rho = 0,\ldots, m - 1.
\end{equation} 

Since $L$ is an homogeneous function, the Euler theorem gives, taking eq.\ (\ref{Dyn4}) into account,
\begin{equation}
L + p_{\alpha} b^{\alpha} = 0
\end{equation}
and by differentiation we obtain
\begin{equation} \label{Diff}
b^{\alpha} dp_{\alpha} + A_{\alpha} L \, \omega^{\alpha}  = 0.
\end{equation}
The differentials $dp_{\alpha}$ and the forms $\omega^{\alpha}$ are arbitrary, apart from the constraints 
\begin{equation} \label{DiffConstr} 
\frac{\partial \Phi_{\rho}}{\partial p_{\alpha}} dp_{\alpha} + 
A_{\alpha} \Phi_{\rho} \, \omega^{\alpha} = 0, \qquad 
\rho = 0,\ldots, m - 1
\end{equation}
and from eq.\ (\ref{Diff}) we have
\begin{equation}  \label{Dyn5}
b^{\alpha} = \sum_{\rho} \alpha^{\rho} \frac{\partial \Phi_{\rho}}{\partial p_{\alpha}}, 
\end{equation}
\begin{equation}  \label{Deri}
A_{\alpha} L = \sum_{\rho} \alpha^{\rho} A_{\alpha}\Phi_{\rho}. 
\end{equation}
Eq.\ (\ref{Dyn5}) inverts, as far as possible, eq.\ (\ref{Dyn4}). At this level, the functions $\alpha^{\rho}(\tau)$ are arbitrary and, if they are not determined by the dynamical equations, they parametrize the gauge transformations of the system.

If, according to the dynamical equations (\ref{Dyn3}), the quantities $\Phi_{\rho}$ are not conserved, the conditions $\dot \Phi_{\rho} = 0$ determine partially the functions $\alpha^{\rho}$ or give rise to secondary constraints. Some simplification can be obtained from the following consequence of the equations derived above
\begin{eqnarray} \label{Conserv}
&\sum_{\rho} \alpha^{\rho} \dot \Phi_{\rho} = 
\sum_{\rho} \alpha^{\rho} \left( \frac{\partial \Phi_{\rho}}{\partial p_{\alpha}} \dot p_{\alpha}  + A_{\alpha}\Phi_{\rho} b^{\alpha}\right) =& 
\nonumber \\  
&b^{\alpha} (\dot p_{\alpha} + A_{\alpha} L) = 
b^{\alpha} b^{\beta} p_{\gamma} F_{\beta \alpha}^{\gamma} = 0.&
\end{eqnarray}
It follows that if $\alpha^0 \neq 0$ and the secondary constraints $\dot \Phi_{\rho} = 0$ for $\rho = 1,\ldots, m-1$ are satisfied, $\Phi_0$ is conserved. In particular, if there is only one primary constraint, it is conserved and there are no secondary constraints.

\subsection{Einstein-Maxwell fields.}

Now we consider a less general situation in which the test particle moves in an Einstein gravitational field and a Maxwell electromagnetic field. In order to describe the electromagnetic field, we adopt the procedure indicated in refs.\ \cite{CSVZ,Toller2}, namely we introduce a principal fibre bundle with structural group $SO^{\uparrow}(1, 3) \times U(1)$ which includes the electromagnetic gauge group. Then the manifold $\mathcal{S}$ has dimension $n = 11$ and we have to introduce a new vector field, that we indicate by $A_{\bullet}$ (in order to avoid a two-digit index $\alpha = 10$), which generates the gauge transformations. Eq.\ (\ref{Deriv}) takes the form
\begin{equation} \label{Deriv1}
X(\tau) = b^{\alpha} A_{\alpha} = 
b^{i} A_{i} + 2^{-1} b^{[ik]} A_{[ik]} + b^{\bullet} A_{\bullet},
\qquad b^{[ik]} = - b^{[ki]}.
\end{equation}

The structure coefficients $F_{[ik] \beta}^{\alpha}$ coincide with the structure constants of the Poincar\'e algebra. They can be written in the form
\begin{eqnarray} \label{Structure}
&F_{[ik][jl]}^{[mn]} = \delta_i^m g_{kj} \delta_l^n - \delta_k^m g_{ij} \delta_l^n
- \delta_i^m g_{kl} \delta_j^n + \delta_k^m g_{il} \delta_j^n& \nonumber \\
&- \delta_i^n g_{kj} \delta_l^m + \delta_k^n g_{ij} \delta_l^m
+ \delta_i^n g_{kl} \delta_j^m - \delta_k^n g_{il} \delta_j^m,&
\end{eqnarray}
\begin{equation} \label{Structure1}
F_{[ik] j}^l = \delta_i^l g_{kj} - \delta_k^l g_{ij},
\end{equation}
\begin{equation} \label{Structure2}
F_{[ik][jl]}^m = 0, \qquad F_{[ik] j}^{[mn]} = 0.  
\end{equation} 
We also assume that the torsion $F_{ik}^j$ vanishes.

The structure coefficients $F_{ik}^{\bullet} = F_{ik}$ represent the electromagnetic field strength and $F_{\alpha \bullet}^{\beta} = 0$. The electromagnetic interaction Lagrangian is 
\begin{equation}
L_I = e b^{\bullet}, 
\end{equation} 
where $e$ is the electric charge. We always assume that it is added  to the free Lagrangian $L$.

By means of these equations, we can write eq.\ (\ref{Dyn3}) in the more explicit form
\begin{equation} \label{Dyn6}
\dot p_i = - b_i{}^k p_k + F_i, \quad 
F_i =  b^k G_{ik} - A_i L, \quad
G_{ik} = 2^{-1} p_{[jl]} F_{ki}^{[jl]} + e F_{ik},
\end{equation}
\begin{equation} \label{Dyn7}
\dot p_{[ik]} = b_i p_k - b_k p_i - b_i{}^j p_{jk} - b_k{}^j p_{ij} - A_{[ik]} L,
\end{equation}
\begin{equation}
\dot p_{\bullet} = 0, \qquad p_{\bullet} = - e,
\end{equation}
where $F_{ki}^{[jl]}$ is the Riemann curvature tensor which describes the gravitational field.
The quantities $p^i$ and $-p_{ik}$ are interpreted as the energy, the momentum and the relativistic angular momentum of the particle, measured in the local reference frame $s$. The quantity $-p_{\bullet}$ is the conserved electric charge.

It is sometimes convenient to use a three-dimensional vector formalism. We introduce the vectors 
\begin{equation} \label{Vector1}
\mathbf{b} =(b^1, b^2, b^3), \qquad
\mathbf{b}' =(b^{[23]}, b^{[31]}, b^{[12]}), \qquad
\mathbf{b}'' =(b^{[10]}, b^{[20]}, b^{[30]}),
\end{equation}
\begin{equation} \label{Vector2}
\mathbf{p} = - (p_1, p_2, p_3), \quad
\mathbf{p}' = - (p_{[23]}, p_{[31]}, p_{[12]}), \quad
\mathbf{p}'' = - (p_{[10]}, p_{[20]}, p_{[30]}),
\end{equation}
\begin{equation} 
\mathbf{f} = - (b^0)^{-1} (F_1, F_2, F_3)
\end{equation}
and we obtain, if $L$ does not depend on $s$,
\begin{equation} \label{PDot1}
\dot p_0 = - \mathbf{b}'' \cdot \mathbf{p} + \mathbf{b} \cdot \mathbf{f},
\end{equation}
\begin{equation}  \label{PDot2}
\dot \mathbf{p} = - \mathbf{b}' \times \mathbf{p} - p_0 \mathbf{b}'' + b^0 \mathbf{f},
\end{equation}
\begin{equation} \label{PDot3}
\dot \mathbf{p}' = - \mathbf{b} \times \mathbf{p}
- \mathbf{b}' \times \mathbf{p}' - \mathbf{b}'' \times \mathbf{p}'',
\end{equation} 
\begin{equation} \label{PDot4}
\dot \mathbf{p}'' = p_0 \mathbf{b} - b^0 \mathbf{p}
- \mathbf{b}' \times \mathbf{p}'' + \mathbf{b}'' \times \mathbf{p}'.
\end{equation}

If we disregard the term $A_i L$ and we put
\begin{equation} \label{EB}
\hat \mathbf{E} = (G_{01}, G_{02}, G_{03}), \qquad 
\hat \mathbf{B} = (G_{32}, G_{13}, G_{21}),
\end{equation}
we can write
\begin{equation} \label{EBf}
\mathbf{f} = \hat \mathbf{E} + (b^0)^{-1} \mathbf{b} \times \hat \mathbf{B}.
\end{equation}
If there is no gravitational field, we have
\begin{equation}
\hat \mathbf{E} = e \mathbf{E}, \qquad \hat \mathbf{B} = e \mathbf{B}
\end{equation}
and the force $\mathbf{f}$ is given by the usual Lorentz formula

\subsection{Noether's theorem.}

In order to treat the connection between symmetries and conservation laws (Noether's theorem), we consider a vector field
\begin{equation} \label{B1}
Y(s) = a^{\alpha}(s) A_{\alpha}(s)
\end{equation}
and the corresponding one-parameter diffeomorphism group $\exp(\epsilon Y)$, which transforms the trajectory $s(\tau)$ into the trajectories $s(\epsilon, \tau)$. If this transformation does not change the action, the expression (\ref{Variation}) vanishes and, taking the dynamical equations into account, we obtain the conservation law 
\begin{equation} \label{Noether}
a^{\alpha} p_{\alpha} = \rm{constant}.
\end{equation} 

We assume that $L$ does not depend on $s$ and consider two applications of this general theorem, which use two different kinds of symmetry properties. In the first case we require
\begin{equation} \label{B2}
[Y, A_{\beta}] = \left( a^{\alpha} F_{\alpha \beta}^{\gamma} - 
A_{\beta} a^{\gamma} \right) A_{\gamma} = 0
\end{equation}
and from eq.\ (\ref{DeltaB}) we see that $\delta b^{\alpha} = 0$, and therefore $\delta L = 0$.  Note that the validity of the conservation law (\ref{Noether}) does not depend on the form of the function $L(b^{\alpha})$.

In the absence of gravitational and electromagnetic fields, $\mathcal{S}$ is the bundle of the Lorentz frames of the Minkowski space-time and the orthochronous Poincar\'e group $\mathcal{P}$ acts freely and transitively on $\mathcal{S}$. We may choose a fixed frame $\hat s$ and represent univocally all the other frames in the form $s = g\hat s$ with $g \in \mathcal{P}$. We may identify $\mathcal{S}$ and $\mathcal{P}$. The vector fields $A_{\alpha}$ generate the left translations of $\mathcal{P}$, but one can also introduce the vector fields $B_{\alpha}$, which generate the right translations, interpreted as Poincar\'e transformations of the fixed frame $\hat s$. They commute with $A_{\alpha}$ and are given by
\begin{equation} 
Y = B_{\alpha}(g) = S_{\alpha}^{\beta}(g) A_{\beta}(g),
\end{equation}
where $S(g)$ is the adjoint representation of $\mathcal{P}$. It follows from the Noether theorem that the quantities
\begin{equation}
\hat p_{\alpha} = S_{\alpha}^{\beta}(g(\tau)) p_{\beta}(\tau)
\end{equation}
are conserved. They are the energy-momentum and the relativistic angular momentum measured in the fixed frame $\hat s$.

In a second application of Noether's theorem we assume eqs.\ (\ref{Structure}--\ref{Structure2}) and consider the infinitesimal Lorentz transformation generated by $A_{[ik]}$. From eq.\ (\ref{DeltaB}) we obtain 
\begin{equation}
\delta b^{\alpha} = - \epsilon F_{[ik] \gamma}^{\alpha} b^{\gamma},
\end{equation}
namely an infinitesimal Lorentz transformation of the quantities $b^{\alpha}$. If the Lagrangian is invariant under this transformation, the quantity $p_{[ik]}$ is conserved. In particular, if the Lagrangian is a Lorentz scalar function of $b^{\alpha}$ all the six quantities $p_{[ik]}$ are conserved. If the Lagrangian is only a rotational scalar, only the three quantities $p_{[rs]}$ with $r, s = 1, 2, 3$ are conserved.  These conservation laws depend on the invariance properties of the Lagrangian and on the form of the structure coefficients $F_{[ik] \gamma}^{\alpha}$. They are also valid in the presence of gravitational and electromagnetic fields.

\section{The presymplectic formalism.}

As it is explained in refs.\ \cite{Souriau} and \cite{Kunzle}, the presymplectic formalism is a valid alternative to the Lagrangian and Hamiltonian formalisms, expecially in dealing with relativistic and constrained systems, and it provides a natural starting point for quantization. In the present section we summarize the formalism and we compare it with the Lagrangian formalism of section 3.

A presymplectic system is given by a $N$-dimensional manifold $\mathcal{E}$ called the {\it evolution space} and a {\it presymplectic form} $\Omega$, namely a degenerate closed differential two-form with constant rank $N - d$.  Note that the rank must be even. One can introduce in $\mathcal{E}$ a time coordinate, but it is not uniquely defined. The kernel of $\Omega$ has dimension $d > 0$ and it is composed of all the vectors $X$ with the property
\begin{equation} \label{Kernel}
i_{X} \Omega = 0,
\end{equation}
where $i_{X}$ is the inner product operator acting on differential forms. In the following $X$ means a vector field satisfying this condition at all the points of $\mathcal{E}$.

The set of vector fields defined by eq.\ (\ref{Kernel}) satisfies the condition of the Frobenius theorem \cite{KN} and it defines a {\it foliation} of $\mathcal{E}$. This means that for every point $x \in \mathcal{E}$ there are submanifolds with dimension $d$ containing it and tangent to all the vector fields $X$. One of these submanifolds contains all the others and is called the {\it maximal leaf} containing $x$. If $d = 1$ the leaves are lines which describe the motion of a representative point in $\mathcal{E}$. These lines, with a suitable parametrization, are also described by the function $\tau \to x(\tau) \in \mathcal{E}$ satisfying the differential equation
\begin{equation}
\dot x(\tau) = X(x(\tau)).
\end{equation}
If $d > 1$, the line that describes the motion is contained in a $d$-dimensional leaf, but it is not uniquely determined by the intial conditions, because there are gauge symmetries.

Under suitable conditions, the set of all the maximal leaves is a symplectic manifold with dimension $N - d$, called the {\it space of motions} \cite{Souriau}. It corresponds to the phase space of the usual Hamiltionian formalism. One can define in this manifold the Poisson brackets and use them for quantization.

In the models we shall consider, the evolution space $\mathcal{E}$ is a submanifold of the cotangent bundle $T^*\mathcal{S} = \mathcal{T}^* \times \mathcal{S}$. Its points, as all the points of $T^*\mathcal{S}$, are indicated by $x = (p_{\alpha}, s)$, $s \in \mathcal{S}$ and satisfy $m$ constraints of the kind (\ref{Constraints}). 
We indicate by the same symbol the differential forms $\omega^{\beta}$ on the space $\mathcal{S}$ and their pull-backs in the space $\mathcal{E}$. Their exterior derivatives are given by
\begin{equation}
d \omega^{\gamma} = - 2^{-1} F_{\alpha \beta}^{\gamma} \, \omega^{\alpha} \wedge \omega^{\beta}.
\end{equation}

The presymplectic form is given by
\begin{equation} \label{Pre}
\Omega = \Omega_0 - d (p_{\alpha} \omega^{\alpha})
= \Omega_0 - dp_{\alpha} \wedge \omega^{\alpha} + 2^{-1} p_{\gamma} F_{\alpha \beta}^{\gamma} \, \omega^{\alpha} \wedge \omega^{\beta},
\end{equation}
where $\Omega_0$ is a closed two-form of the kind
\begin{equation} 
\Omega_0 = 2^{-1} C^{\alpha \beta}(p) \, d p_{\alpha} \wedge d p_{\beta}.
\end{equation}
In this article we consider only models with $\Omega_0 = 0$.

If we put
\begin{equation} 
i_X dp_{\alpha} = X p_{\alpha} = \dot p_{\alpha}, \qquad
i_X \omega^{\alpha} = b^{\alpha},
\end{equation}
we have
\begin{equation} 
0 = i_X \Omega = - \dot p_{\beta} \omega^{\beta} + b^{\beta} dp_{\beta} 
+ p_{\gamma} b^{\alpha} F_{\alpha \beta}^{\gamma} \omega^{\beta}.
\end{equation}
Since the differential forms $dp_{\alpha}$ and $\omega^{\alpha}$ are arbitrary, apart from the constraints (\ref{DiffConstr}), we obtain eq.\ (\ref{Dyn5}) and
\begin{equation} \label{Dyn8}
\dot p_{\alpha} = 
b^{\gamma} p_{\beta} F_{\gamma \alpha}^{\beta} - \sum_{\rho} \alpha^{\rho} A_{\alpha}\Phi_{\rho},
\end{equation} 
If we take eq.\ (\ref{Deri}) into account, we find that the last equation is equivalent to eq.\ (\ref{Dyn3}). We see that the presymplectic formalism includes the Lagrangian formalism of section 3.2. The advantages of the presymplectic formalism are that one can choose more freely the constraints and one can build more general models with $\Omega_0 \neq 0$ (see ref.\ \cite{Vanzo}). Note that eq.\ (\ref{Conserv}) is also generally valid within the presymplectic formalism.

In the practical applications,  by requiring the consistency of the primary constraints with the dynamical equations, we may find conditions on the functions $\alpha^{\rho}$ or secondary constraints. We use this term also for tertiary and further constraints. Their existence means that the form $\Omega$ was not degenerate at all the points of $\mathcal{E}$, but only in a submanifold $\mathcal{E}'$ defined by the secondary constraints. The manifold $\mathcal{E}$ was not a true presymplectic evolution space and one may try to promote the secondary constraints to the role of primary constraints, namely to consider the evolution space $\mathcal{E'}$, with the hope that it has the required properties (that is not evident). Then we have to add new terms to the sum in eq.\ (\ref{Dyn8}) and to introduce new variables  $\alpha^{\rho}$, which, if they are not fixed by the dynamical equations, describe new gauge transformations. In the following, we shall not use this delicate procedure \cite{Dirac}, which is, however, a necessary step for the construction of a symplectic phase space and for the quantization.

In in a presymplectic formalism, a function $q(p_{\alpha}, s)$ defined on $\mathcal{E}$ is conserved if it is constant on the leaves, namely if $X q = 0$ for all the vector fields $X$ which satisfy eq.\ (\ref{Kernel}). In order to formulate the Noether theorem, we consider a vector field $Y$ defined on $\mathcal{E}$ which generates a group of transformations which leave the presymplectic form $\Omega$ invariant. It has the property
\begin{equation} 
d i_Y \Omega = 0
\end{equation}
and then one can find, at least locally, a function $q$ with the property
\begin{equation} 
i_Y \Omega = - d q.
\end{equation}
If $X$ satisfies eq.\ (\ref{Kernel}), we have
\begin{equation} 
X q = i_X dq = - i_X i_Y \Omega = i_Y i_X \Omega = 0
\end{equation}
and the quantity $q$ is conserved.

Since there is no danger of confusion, we use the symbol $A_{\alpha}$ to indicate a vector field defined in $T^*\mathcal{S}$ in such a way that $A_{\alpha} p_{\beta} =0$ and its projection on $\mathcal{S}$ is the vector field $A_{\alpha}$ introduced in section 3.1. If the constraints do not depend on $s$, these fields are tangent to the submanifold $\mathcal{E}$  and we can consider them as vector field defined on it.

As a first application of this version of the Noether theorem, we consider the vector field (\ref{B1}) satisfying eq.\ (\ref{B2}) and extended to $\mathcal{E}$ in the way explained above. We have
\begin{equation} 
i_Y \Omega = a^{\alpha} dp_{\alpha} +  p_{\gamma} F_{\alpha \beta}^{\gamma} \, a^{\alpha} \omega^{\beta} =  d (a^{\alpha} p_{\alpha}).
\end{equation}
In the last step we have used eq.\ (\ref{B2}). We see that the quantity (\ref{Noether}) is conserved, exactly as in the Lagrangian formalism of section 3.4.

In a second application, we assume that the structure coefficients $F_{[ik] \alpha}^{\beta}$ are given by eqs.\ (\ref{Structure}--\ref{Structure2}) and we consider the vector field
\begin{equation} 
Y = A_{[ik]} + F_{[ik] \alpha}^{\beta} p_{\beta} \frac{\partial}{\partial p_{\alpha}},
\end{equation}
which generates an infinitesimal Lorentz transformation in $T^*\mathcal{S}$. If the submanifold $\mathcal{E}$ is invariant under this transformation, $Y$ can be considered as a vector field on $\mathcal{E}$. With some calculations we obtain
\begin{equation} 
i_Y \Omega = d p_{[ik]}.
\end{equation}
and $p_{[ik]}$ is conserved, in analogy with the results of section 3.4.

\section{Frames and particles.}

\subsection{Rest frames and acceleration.}

In order to interpret the solutions of the dynamical equations, we have to consider the connection between the Lorentz frames $s(\tau)$ and the physical particle. We indicate by $\pi: \mathcal{S} \to \mathcal{M}$ the projection which associates to the tetrad $s \in \mathcal{S}$ its origin $x = \pi(s) \in \mathcal{M}$. The projection of the curve $\tau \to s(\tau)$ is the world line  $\tau \to \pi(s(\tau)) = x(\tau)$ in the the spacetime manifold $\mathcal{M}$. From eq.\  (\ref{Deriv}), remembering that the vector fields $A_i$ generate the parallel displacements along the four-vectors $e_i(\tau)$ of the tetrad $s(\tau)$, we have
\begin{equation}
\dot x(\tau) = \frac{dx(\tau)}{d \tau} = b^i(\tau) e_i(\tau).
\end{equation}
If $\tau$ is the proper time, namely if $b \cdot b = 1$, the quantities $b^i$ are the compoments of the velocity of the origin of $s(\tau)$, with respect to the same  frame.

The tetrad $s(\tau + d\tau)$ differs from the parallel transported tetrad by an infinitesimal Lorentz transformation with parameters $b^{[ik]} d \tau$. This means that the covariant derivatives of the tetrad four-vectors are given by
\begin{equation}
\frac{D e_i(\tau)}{d \tau} = - b_i{}^k(\tau) e_k(\tau),
\end{equation}
and we obtain the formula
\begin{equation}
a(\tau) = \frac{D \dot x(\tau)}{d \tau} = \dot b^i  e_i -  b^i b_i{}^k e_k =
(\dot b^i + b^i{}_k b^k) e_i. 
\end{equation}

If $\tau$ is the proper time, $a$ is the covariant acceleration four-vector and its components in the moving frame $s(\tau)$ are given by
\begin{equation}
a^i = \dot b^i + b^i{}_k b^k. 
\end{equation}

A {\it rest frame} is defined by the condition $\mathbf{b} = 0$ and the four-velocity of the origin of the frame is equal to the tetrad vector $e_0$, which is timelike by definition. If $\tau$ is the proper time, we also have $b^0 = 1$, and $d \tau$ is a time measured in the rest frame. We have
\begin{equation}
a = \frac{D e_0}{d \tau} =  b^{r0} e_r, \qquad  a^0 = 0, \qquad \mathbf{a} = \mathbf{b}''.
\end{equation}
The vector $\mathbf{a}$ represents the acceleration of the origin of a rest frame measured in the same frame. If the parameter $\tau$ is arbitrary, we can write the more general formula
\begin{equation} \label{Acc}
\mathbf{a} = (b^0)^{-1} \mathbf{b}''.
\end{equation}

\subsection{Zero-momentum frames and pseudo-acceleration.}

The condition $\mathbf{p} = 0$ defines a {\it zero-momentum frame}. This condition implies that the energy-momentum is a timelike four-vector. In the case of a free particle, the center of mass is at rest in this frame, but this is not true in general. In a zero-momentum frame $\mathbf{p}'$ represents the spin of the particle, as we see considering the Pauli-Lubanski four-vector (\ref{PL}). One can still consider the vector (\ref{Acc}) but it does not represent the acceleration of the origin $x(\tau)$ and not even the acceleration of the center of mass, as we show below. We call it the {\it pseudo-acceleration}. From eq.\ (\ref{PDot2}) we have
\begin{equation} 
\mathbf{a} = (p_0)^{-1} \mathbf{f}.
\end{equation}
Note that $p_0$ in a zero-momentum frame is the invariant mass and this formula coincides with the Newton formula, valid for a point particle. If the particle is extended, this formula remains valid if we replace the acceleration by the pseudo-acceleration. This may be considered as the physical meaning of the pseudo-acceleration.

\subsection{Center of mass and central frames.}

In order to describe the position of the center of mass, we have to introduce for each tetrad a system of coordinates in a suitable open set of $\mathcal{M}$, for instance a system of normal coordinates (see, for instance, \cite{Toller2,MT}). In the absence of gravitation, namely in special relativity theory,  one can associate to every local frame $s$ a Lorentzian coordinate system in the flat Minkowski spacetime and, according to eq.\ (\ref{COM}), the space coordinates of the center of mass at zero time in this frame can be written in the form
\begin{equation} \label{COM2}
\mathbf{y} = - (p_0)^{-1} \mathbf{p}''.
\end{equation}
The condition $\mathbf{p}'' = 0$ means that the trajectory of the center of mass crosses the origin. A frame with this property is called a {\it central frame}. This definition can be extended to the case in which gravitation is present. From eq.\ (\ref{PDot4}) we see that, if the spin is zero, a zero-momentum central frame is also a rest frame.

The definition $\mathbf{p}'' = 0$ is not Lorentz invariant and it is often replaced by the condition
\begin{equation} \label{Dixon}
p_{ik} p^k = 0,
\end{equation}
proposed by Dixon \cite{Dixon}. A frame with this property is called a {\it Dixon frame}. It can always be obtained from a central zero-momentum frame by means of a Lorentz transformation.

In a fixed reference frame, the four-velocity of the center of mass is parallel to the energy-momentum four-vector. One cannot deduce, in general, that in a zero-momentum moving frame the four-vector $e_0$ is the four-velocity of the center of mass, because the definiton of the center of mass may depend on the velocity of the frame \cite{Moller}. Moreover, the parameter $\tau$ defined by the condition $b^0 = 1$ is not the proper time of the center of mass. It follows that the quantity (\ref{Acc}), is not directly related to the acceleration of the center of mass.

The models treated in the following sections concern extended particles which contain a point charge. In this case, instead of working with central or Dixon frames, it is convenient to assume that the charge lies at the origin of $s(\tau)$. More in general, if the charge is not pointlike, one can require that the electric dipole moment with repect to the frame $s(\tau)$ vanishes.

\subsection{A classical model.}

In order to permit an easier comparison with the models treated in the following sections, we summarize, using our notations, the presymplectic treatment, given in ref.\ \cite{Kunzle}, of a massive particle with spin and a magnetic dipole moment.
A Lagrangian treatment of the same problem is given, for instance, in refs.\ \cite{CSVZ2,Preti}. We  start from the constraints
\begin{equation} 
\Phi_0 = p_0 - m + \boldsymbol{\mu} \cdot \mathbf{B} = 0, \qquad \mathbf{p}' = \boldsymbol{\sigma}, \qquad \mathbf{p} = \mathbf{p}'' = 0,
\end{equation}
where $m$ is the mass, $\boldsymbol{\sigma}$ is a fixed vector describing the spin and $\boldsymbol{\mu}$ is the magnetic dipole moment. $s(\tau)$ is a central zero-momentum frame.  The scalar constraint coincides, taking into accout that $\mathbf{p} = 0$, with the constraint derived from the Lagrangian treatment of ref.\ \cite{Preti}. In order to avoid the introduction of new degrees of freedom, we assume that the magnetic dipole moment is proportional to the spin, namely we put
\begin{equation} 
\boldsymbol{\mu}  = g e (2m)^{-1} \boldsymbol{\sigma}.   
\end{equation}
For a Dirac particle we have $g = 2$.

The scalar constraint $\Phi_0$ depends on $s$ through the magnetic field $\mathbf{B}$ and the last term in eq.\ (\ref{Dyn8}) has to be taken into account. Since $F_{ik}$ is a tensor field, its infinitesimal Lorentz transformations are given by
\begin{equation} 
A_{ik} F_{jl} = g_{kj} F_{il} - g_{ij} F_{kl} - g_{kl} F_{ij} + g_{il} F_{kj}, 
\end{equation}
while $A_i F_{jl}$ is its covariant derivative. After some calculation, we see that  eqs.\ (\ref{PDot1}--\ref{PDot4}) with the new term take the form
\begin{equation} \label{CPDot1}
\dot p_0 = \mathbf{b} \cdot \hat \mathbf{E}
- b^0 A_0 (\boldsymbol{\mu} \cdot \mathbf{B}),
\end{equation}
\begin{equation} \label{CPDot2}
p_0 \mathbf{b}'' = b^0 \hat \mathbf{E} + \mathbf{b} \times \hat \mathbf{B}
+ b^0 \mathbf{A} (\boldsymbol{\mu} \cdot \mathbf{B}), 
\end{equation}
\begin{equation} \label{CPDot3}
\mathbf{b}' \times \boldsymbol{\sigma} =
b^0 \boldsymbol{\mu} \times \mathbf{B},
\end{equation} 
\begin{equation} \label{CPDot4}
p_0 \mathbf{b} = -\mathbf{b}'' \times \boldsymbol{\sigma}
- b^0 \boldsymbol{\mu} \times \mathbf{E}.
\end{equation}
We have used the vector notation $\mathbf{A} = (A_1, A_2, A_3)$.

Note that $b^0$ and the component of $\mathbf{b}'$ parallel to $\boldsymbol{\sigma}$ are not determined, namely there is a gauge symmetry. We can fix the gauge by assuming $b^0 = 1$ and, instead of eq.\ (\ref{CPDot3}), the stronger equation
\begin{equation} \label{CPDot5}
\mathbf{b}' = - g e (2m)^{-1} b^0 \mathbf{B}.
\end{equation} 
It gives the angular velocity $\mathbf{b}'$ of the precession of the moving frame and of the spin with respect to a Fermi-Walker transported zero-momentum frame.

Eq.\ (\ref{CPDot1}) is a consequence of the other equation and we use eqs.\ (\ref{CPDot2}) and  (\ref{CPDot4}) to determine $\mathbf{b}$ and $\mathbf{b}''$.  We obtain
\begin{eqnarray} 
&\mathbf{b}'' = ((p_0)^2 - \boldsymbol{\sigma} \cdot \hat \mathbf{B})^{-1}
\left (p_0 (\hat \mathbf{E} + \mathbf{A} 
(\boldsymbol{\mu} \cdot \mathbf{B}))\right.& 
\nonumber \\ 
&\left. - (p_0)^{-1} (\hat \mathbf{E} \cdot \hat \mathbf{B}  
+ \hat \mathbf{B} \cdot \mathbf{A} (\boldsymbol{\mu} \cdot \mathbf{B})) 
\boldsymbol{\sigma} - (\boldsymbol{\mu} \cdot \hat \mathbf{B}) \mathbf{E}
+ (\mathbf{E} \cdot \hat \mathbf{B}) \boldsymbol{\mu} \right).&
\end{eqnarray}
We see that, as it is remarked in ref.\ \cite{Kunzle}, to avoid singularities one has to assume
\begin{equation} \label{Sing}
p_0 > 0, \qquad (p_0)^2 \neq \boldsymbol{\sigma} \cdot \hat \mathbf{B}.
\end{equation}

Since the ten quantities $p^{\alpha}$ are fixed by the constraints, the initial conditions are given by the three space coordinates of the frame origin and five parameters for the orientation of the tetrad, taking into account that the rotations around $\boldsymbol{\sigma}$ are gauge transformations.  The phase space has dimension eight, as it is expected for a massive spinning particle \cite{Souriau}.
 
\section{A highly symmetric Lagrangian.}

The quantities $b^{\alpha}$ describe the velocity, the acceleration and the angular velocity of the Lorentz frame $s(\tau)$ and, as we have seen in the preceding section, are related to the quantities that describe the motion of the particle. Limitations to the values of these quantities can be described by the condition
\begin{equation} \label{Cond3}
X(\tau) = b^{\alpha} A_{\alpha} \in \mathcal{T}^+ \subset \mathcal{T},
\end{equation}
where $\mathcal{T}^+$ is the cone found in refs.\ \cite{Toller3,Toller4}, starting from some simple and natural conditions, which determine it up to the choice of the numerical value $a_M = \lambda^{-1}$ of the maximal acceleration.

A simple definition of $\mathcal{T}^+$ is obtained by introducing the real symmetric $4 \times 4$ matrix
\begin{equation} \label{MatrixB}
\hat b = -i b^i C^{-1} \gamma_i
+ 2^{-1} \lambda b^{[ik]} C^{-1} \gamma_{i} \gamma_{k},
\end{equation} 
where $\gamma_{i}$ are the pure imaginary Dirac matrices in the Majorana representation. The real antisymmetric matrix $C$ has the property
\begin{equation} 
\gamma_{i}^T = - C^{-1} \gamma_{i} C.
\end{equation}
One can choose a representation in which $C = -i \gamma_0$, but one has to remember that the $\gamma$ matrices represent linear transformations in the spinor space, while $C^{-1}$ represents an antisymmetric bilinear form.

One defines $\mathcal{T}^+$ by requiring that $\hat b$ is positive semidefinite, namely that
\begin{equation} \label{Cond8}
\psi^T \hat b \psi \geq 0
\end{equation}
for any choice of the real spinor $\psi$. Note that if we describe the electromagnetic field introducing an 11-dimensional manifold $\mathcal{S}$, the quantity $b^{\bullet}$ does not appear in the definition of $\mathcal{T}^+$, which should be called more correctly a wedge. It is clear that $\mathcal{T}^+$ is symmetric under the transformations
\begin{equation} \label{Symmetry}
\hat b \to (a^{-1})^T \hat b a^{-1}, \qquad a \in GL(4, \mathbf{R}).
\end{equation}
The 16-dimensional symmetry group $GL(4, \mathbf{R})$ contains a subgroup isomorphic to $SL(2, \mathbf{C})$, namely to the universal covering of the proper orthochronous Lorentz group. One shows \cite{Toller3,Toller4} that $X \in \mathcal{T}^+$ implies the inequalities
\begin{equation} \label{Bounds}
\|\mathbf{b}\| \leq b^0, \qquad \|\mathbf{b}'\| \leq \lambda^{-1} b^0, \qquad 
\|\mathbf{b}''\| \leq \lambda^{-1} b^0, 
\end{equation}
which can be interpreted as bounds on the velocity, the angular velocity and the acceleration of the frame $s(\tau)$.

On the boundary of $\mathcal{T}^+$ we have $\det \hat b = 0$ and from the rule proposed in the introduction we obtain the free Lagrangian
\begin{equation}  \label{Lagr4}
L(X) = - m (\det \hat b)^{1/4}.
\end{equation}
It has the large symmetric group $SL(4, \mathbf{R})$, which is broken by the presence of the structure coefficients $F_{\alpha \beta}^{\gamma}$, which have a lower symmetry. We shall see that this Lagrangian does not provide a model of the kind we are looking for, but it gives the occasion for some importent developments that will be useful in the following sections.

A direct computation \cite{Toller4} gives
\begin{eqnarray} \label{Lagr5}
&L = -m \left(((b^0)^2 - \|\mathbf{b}\|^2 - \lambda^2 \|\mathbf{b}'\|^2 - 
\lambda^2 \|\mathbf{b}''\|^2 )^2 - 4 \lambda^2 \|\mathbf{b} \times \mathbf{b}'\|^2
\right.& \nonumber \\
&\left. - 4 \lambda^4 \|\mathbf{b}' \times \mathbf{b}''\|^2
- 4 \lambda^2 \|\mathbf{b}'' \times \mathbf{b}\|^2  
+ 8 \lambda^2 b^0 \, \mathbf{b} \cdot \mathbf{b}' \times \mathbf{b}''\right)^{1/4}&
\end{eqnarray}
and the momenta (\ref{Dyn4}) have the form
\begin{eqnarray} \label{P1}
&p_0 = m^4 |L|^{-3} \left(((b^0)^2 - \|\mathbf{b}\|^2 - \lambda^2 \|\mathbf{b}'\|^2 - \lambda^2 \|\mathbf{b}''\|^2) b^0 \right.& \nonumber \\ 
&\left. + 2 \lambda^2 \mathbf{b} \cdot \mathbf{b}' \times \mathbf{b}'' \right),&
\end{eqnarray}
\begin{eqnarray} \label{P2}
&\mathbf{p} = m^4 |L|^{-3} \left(((b^0)^2 - \|\mathbf{b}\|^2  
+ \lambda^2 \|\mathbf{b}'\|^2  
+ \lambda^2 \|\mathbf{b}''\|^2) \mathbf{b} \right.& \nonumber \\
&\left. - 2 \lambda^2 (\mathbf{b} \cdot \mathbf{b}') \mathbf{b}' -
2\lambda^2 (\mathbf{b} \cdot \mathbf{b}'') \mathbf{b}'' - 
2 \lambda^2 b^0 \, \mathbf{b}' \times \mathbf{b}'' \right),&
\end{eqnarray}
\begin{eqnarray} \label{P3}
&\mathbf{p}' = m^4 |L|^{-3} \lambda^2 \left(((b^0)^2 + \|\mathbf{b}\|^2  
- \lambda^2 \|\mathbf{b}'\|^2 
+ \lambda^2 \|\mathbf{b}''\|^2) \mathbf{b}' \right.& \nonumber \\
&\left. - 2(\mathbf{b'} \cdot \mathbf{b}) \mathbf{b}
- 2\lambda^2 (\mathbf{b'} \cdot \mathbf{b}'') \mathbf{b}'' 
- 2 b^0 \, \mathbf{b}'' \times \mathbf{b} \right),&
\end{eqnarray}
\begin{eqnarray} \label{P4}
&\mathbf{p}'' = m^4 |L|^{-3} \lambda^2 \left(((b^0)^2 + \|\mathbf{b}\|^2  
+ \lambda^2 \|\mathbf{b}'\|^2  
- \lambda^2 \|\mathbf{b}''\|^2) \mathbf{b}'' \right.& \nonumber \\
&\left. - 2(\mathbf{b''} \cdot \mathbf{b}) \mathbf{b} 
- 2\lambda^2 (\mathbf{b''} \cdot \mathbf{b}') \mathbf{b}' 
- 2 b^0 \, \mathbf{b} \times \mathbf{b}' \right).&
\end{eqnarray}

In order to find the primary constraint, it is convenient to use the spinor formalism. We introduce the matrix
\begin{equation} \label{HatP}
\hat p = i p_i \gamma^i C
- (2 \lambda)^{-1} p_{[ik]} \gamma^{i} \gamma^{k} C,
\end{equation}
and eq.\ (\ref{Dyn4}) takes the form (for a fixed value of $\tau$)
\begin{equation}
- d L = p_{\alpha} d b^{\alpha} = p_k d b^k + 2^{-1} p_{[ik]} d b^{[ik]} = 
2^{-2} \, \mathrm{Tr}(\hat p d \hat b).
\end{equation}
From eq.\ (\ref{Lagr4}), using the formula
\begin{equation}
d \det \hat b = \det \hat b \, \mathrm{Tr} (\hat b^{-1} d \hat b),
\end{equation}
we obtain 
\begin{equation} \label{Mom}
\hat p = m (\det \hat b)^{1/4} \hat b^{-1}.
\end{equation}

It follows that
\begin{equation} \label{Prim}
\Phi_0 = \det \hat p - m^4 = 0.
\end{equation}
This is a primary constraint \cite{Dirac} due to the arbitrariness of the parameter $\tau$. We also obtain the formula
\begin{equation} \label{Inverse}
\hat b = \alpha' \hat p^{-1},
\end{equation}
which determines the matrix $\hat b$ up to a numerical factor $\alpha'$. It follows that there are no additional primary constraints. According to the results of section 3.2, the primary constraint is conserved and there are no secondary constraints.

The manifold defined by the constraint (\ref{Prim}) has three connected components, characterized by the (even) number of negative eigenvalues of the matrix $\hat p$. For reasons of continuity, a trajectory cannot pass from one connected component to another one. On the physically relevant component $\hat p$ has four positive eigenvalues and it is definite positive. Note that also in the usual relativistic formalism one has to choose one of the two connected components of the constraint manifold defined by $p \cdot p = m^2$. From eq.\ (\ref{Inverse}), we see that, if we choose $\alpha' > 0$, the matrix $\hat b$ is positive definite and the condition (\ref{Cond3}) is satisfied for all the values of the parameter $\tau$.

We have obtained a mathematically consistent dynamical system which (at least locally) determines uniquely (up to reparametrization) a trajectory for each initial condition $(s, p_{\alpha})$ which satisfies the primary constraint (\ref{Prim}). We see that the phase space has dimension eighteen. These conclusions are also valid for nonlocal theories, in which the space $\mathcal{S}$ is not the bundle of the Lorentz frames of a space-time manifold and the eqs.\ (\ref{Structure}--\ref{Structure2}) are not valid.

The primary constraint can also be written in the form
\begin{eqnarray} \label{Prim2}
& m^4 = \det \hat p = \left((p_0)^2 - \|\mathbf{p}\|^2 - \lambda^{-2} \|\mathbf{p}'\|^2 - 
\lambda^{-2} \|\mathbf{p}''\|^2 \right)^2 - 
4 \lambda^{-2} \|\mathbf{p} \times \mathbf{p}'\|^2& \nonumber \\
& - 4 \lambda^{-4} \|\mathbf{p}' \times \mathbf{p}''\|^2
- 4 \lambda^{-2} \|\mathbf{p}'' \times \mathbf{p}\|^2 - 
8 \lambda^{-2} p_0 \, \mathbf{p} \cdot \mathbf{p}' \times \mathbf{p}''&
\end{eqnarray}
and from eq.\ (\ref{Dyn5}) we obtain
\begin{eqnarray} \label{Inverse1}
&b^0 = \alpha \left(((p_0)^2 - \|\mathbf{p}\|^2 - \lambda^{-2} \|\mathbf{p}'\|^2 - 
\lambda^{-2} \|\mathbf{p}''\|^2) p_0 \right.& \nonumber \\
&\left. - 2 \lambda^{-2} \mathbf{p} \cdot \mathbf{p}' \times \mathbf{p}'' \right),&
\end{eqnarray}
\begin{eqnarray} \label{Inverse2}
&\mathbf{b} = \alpha \left(((p_0)^2 - \|\mathbf{p}\|^2 + \lambda^{-2} \|\mathbf{p}'\|^2 + 
\lambda^{-2} \|\mathbf{p}''\|^2) \mathbf{p} \right. & \nonumber \\
&\left. - 2\lambda^{-2} (\mathbf{p} \cdot \mathbf{p}') \mathbf{p}' -
2\lambda^{-2} (\mathbf{p} \cdot \mathbf{p}'') \mathbf{p}'' +
 2 \lambda^{-2} p_0 \, \mathbf{p}' \times \mathbf{p}'' \right),&
\end{eqnarray}
\begin{eqnarray} \label{Inverse3}
&\mathbf{b}' = 
\alpha \lambda^{-2} \left(((p_0)^2 + \|\mathbf{p}\|^2 - \lambda^{-2} \|\mathbf{p}'\|^2  
+ \lambda^{-2} \|\mathbf{p}''\|^2) \mathbf{p}' \right. & \nonumber \\
&\left. - 2 (\mathbf{p}' \cdot \mathbf{p}) \mathbf{p} 
- 2 \lambda^{-2} (\mathbf{p}' \cdot \mathbf{p}'') \mathbf{p}'' 
+ 2 p_0 \, \mathbf{p}'' \times \mathbf{p} \right),&
\end{eqnarray}
\begin{eqnarray} \label{Inverse4}
&\mathbf{b}'' =
\alpha \lambda^{-2} \left(((p_0)^2 + \|\mathbf{p}\|^2 + \lambda^{-2} \|\mathbf{p}'\|^2 
- \lambda^{-2} \|\mathbf{p}''\|^2) \mathbf{p}'' \right. & \nonumber \\
&\left. - 2(\mathbf{p}'' \cdot \mathbf{p}) \mathbf{p} 
- 2\lambda^{-2} (\mathbf{p}'' \cdot \mathbf{p}') \mathbf{p}' 
+ 2 p_0 \, \mathbf{p} \times \mathbf{p}' \right),&
\end{eqnarray}
where $\alpha$ is a positive arbitrary function of $\tau$. 

In order to find the energy-momentum spectrum, we remark that 
\begin{equation}
p_0 = 2^{-2} \mathrm{Tr} \hat p \geq m.
\end{equation}
In fact, if we fix the determinant of a positive matrix, its trace is minimal when the matrix is a positive multiple of the identity. This argument also shows that when $p_0 = m$ we must have $\mathbf{p} = \mathbf{p}' = \mathbf{p}'' = 0$. From the Lorentz symmetry it follows that
\begin{equation} \label{Ineq}
p  \cdot p = (p_0)^2 - \|\mathbf{p}\|^2 \geq m^2.
\end{equation}

If the equality holds, we say that the particle lies in a fundamental state and we have $\mathbf{p}' = \mathbf{p}'' = 0$, namely $p_{ik} = 0$. From eqs.\ (\ref{Inverse1}--\ref{Inverse4}) we have
\begin{equation} 
\mathbf{b}' = \mathbf{b}'' = 0, \qquad  (b_0)^{-1} \mathbf{b} = (p_0)^{-1} \mathbf{p}
\end{equation}
We see that the frame $s(\tau)$ is parallel transported and that the four-vectors $b$ and $p$ are parallel.

The quantities $\mathbf{p}'$ and $\mathbf{p}''$ are conserved and from eq.\ (\ref{Dyn6}) we obtain
\begin{equation} 
\dot p_i =  e b^k F_{ik}, 
\end{equation}
namely the usual relativistic formula. In particular, we have
\begin{equation}
\frac{d}{d \tau}(p \cdot p) = 2 p \cdot \dot p = 0.
\end{equation}
This means that the particle remains in a fundamental state, namely it is not excited, even in the presence of gravitational and electromagnetic fields. 

Note that the rotations are not gauge transformations and particles at rest in a fundamental state that differ for a rotation of the frame are different physical states. One may think that the particle is an extended object with no rotational symmetry.

We have seen that the fundamental states describe particles with mass $m$ and spin zero, which in the presence of gravitational and electromagnetic fields follow the usual laws of relativistic dynamics. In particular, there is no upper limit to the acceleration of the center of mass, which coincides with the origin of the moving frame, where the charge is concentrated. Of course, the bounds (\ref{Bounds}) are satisfied, but they do not concern the acceleration of the particle and not even its pseudo-acceleration, because the frame $s(\tau)$ is neither a rest frame nor a zero-momentum frame. We conclude that the model described in the present section does not implement the bound to the acceleration that motivated it and, as we shall see in the following sections, some modification is required.

Besides the fundamental states, there are other solutions. The quantities $\mathbf{p'}$ and $\mathbf{p''}$ do not necessarily vanish, but, since the Lagrangian is a Lorentz scalar, the Noether theorem assures that they are conserved. The four-vectors $p$ and $b$ are not necessarily parallel and the quantity $p \cdot p$ may not be constant, if electromagnetic or gravitational fields are present. This means that the internal degrees of freedom of the particle may be excited.

Since the vectors $\mathbf{p}, \lambda^{-1} \mathbf{p'}, \lambda^{-1} \mathbf{p''}$ appear symmetrically in the constraint equation, besides eq.\ (\ref{Ineq}), we also have the inequalities
\begin{equation} \label{Inequal}
\|\mathbf{p'}\|^2 \leq \lambda^2 ((p_0)^2 -  m^2), \qquad
\|\mathbf{p''}\|^2 \leq \lambda^2 ((p_0)^2 -  m^2).
\end{equation}
The constraint is Lorentz invariant and the same inequalities hold in a Lorentz transformed frame. The first inequality, applied in a zero-momentum frame, gives an upper bound to the spin $\sigma = \|\mathbf{p}'\|$, which, written in terms of Lorentz invariants, takes the form
\begin{equation} \label{SupS}
\sigma^2 \leq \lambda^2 ((p \cdot p) -  m^2), \qquad
\end{equation}
From the second inequality we obtain, the upper bound
\begin{equation} \label{SupY}
\|\mathbf{y}\| < \lambda,
\end{equation}
valid in all the frames, for the distance of the center of mass from the origin.

\section{Models with a maximal acceleration or pseudo-acceleration.}

\subsection{General considerations.}

We have seen in the preceding section that, in order to find models with limitations to the acceleration, one has to introduce additional constraints that limit the choice of the Lorentz frames associated to the particle. We consider two possibilities.
\renewcommand{\labelenumi}{\alph{enumi})}
\begin{enumerate}
\item As it is shown in section 5.1, the acceleration of the ``distinguished point'' of the particle, namely of the origin of $s(\tau)$, is given by eq.\ (\ref{Acc}) if $s(\tau)$ is a rest frame, namely if $\mathbf{b} = 0$. In a Lagrangian model, one can enforce this condition by means of Lagrangian multipliers and in a symplectic model, as we see from eq.\ (\ref{Dyn5}), one has to introduce constraints that do not contain $\mathbf{p}$. In both cases the momentum $\mathbf{p}$ is not subject to any condition and one cannot avoid an unphysical energy-momentum spectrum.
\item One can also require that $s(\tau)$ is a zero-momentum frame, namely that $\mathbf{p} = 0$. This condition is satisfied if the Lagrangian does not depend on $\mathbf{b}$ and can be introduced as an explicit constraint in a presymplectic formalism. In models of this kind the energy-momentum is necessarily a time-like four-vector, but, instead of finding an upper bound for the acceleration, one finds an upper bound for the pseudo-acceleration, defined in section 5.2. The velocity $(b^0)^{-1}\mathbf{b}$ of the origin, namely of the charge, may be larger than the velocity of light, but it does not concern an object that carries energy. The four-velocity of the center of mass (with respect to a fixed frame), which is proportional to the four-momentum, is timelike.
\end{enumerate}

One can introduce further constraint, besides the ones described above. Theories with the constraint $\mathbf{p}'' = 0$, namely theories which consider only central frames, as the one treated in section 5.4, are not interesting from our point of view. In fact their Lagrangian  does not depend on $\mathbf{b}''$, and it is difficult to impose an upper bound to the acceleration or the pseudo-acceleration. It seems that theories with such upper bounds cannot be described by central frames within the framework we are considering.  One may also remark that, for dimensional reasons, a Lagrangian that does not depend on $\mathbf{b}'$ or $\mathbf{b}''$ cannot contain the length $\lambda$. 

Interesting models are obtained by requiring the constraint $\mathbf{p}' = 0$, namely by considering Lagrangians that do not depend on $\mathbf{b}'$. In a zero-momentum frame this means that the Pauli-Lubanski four-vector (\ref{PL}) vanishes and the model describes spinless particles. This is not true, in general, in a rest frame. 

\subsection{Models for spinning particles.}

In order to find a Lagrangian model of the kind b) which describes particles with arbitrary spin, we can consider a Lagrangian depending on $b^0$, $\mathbf{b}'$ and $\mathbf{b}''$. If we think that the model of section 6 contains some relevant physical idea, it is natural to enforce the constraint $\mathbf{p} = 0$ by putting $\mathbf{b} = 0$ in the Lagrangian (\ref{Lagr5}). In this way we obtain
\begin{equation} \label{Lagr8}
L = -m \left(((b^0)^2 - \lambda^2 \|\mathbf{b}'\|^2 - \lambda^2 \|\mathbf{b}''\|^2 )^2 - 
4 \lambda^4 \|\mathbf{b}' \times \mathbf{b}''\|^2 \right)^{1/4}.
\end{equation}

By requiring that the Lagrangian is real we obtain the condition
\begin{equation} \label{Bound}
(b^0)^2 \geq \lambda^2 \left( \|\mathbf{b}'\|^2 + \|\mathbf{b}''\|^2 + 
2 \|\mathbf{b}' \times \mathbf{b}''\| \right),
\end{equation}
which implies the second and the third inequalities of eq.\ (\ref{Bounds}) and gives a joint upper bound to the pseudo-acceleration and the angular velocity. This inequality defines a convex cone $\mathcal{T}_7 \cap \mathcal{T}^+$ in the seven-dimensional vector subspace $\mathcal{T}_7$ of $\mathcal{T}$, defined by $\mathbf{b} = 0$. It is discussed and physically justified by means of a simple geometric model in ref.\ \cite{Toller4}.

Besides the constraint $\mathbf{p} = 0$, we obtain the following expressions for the momenta:
\begin{equation} \label{IP1}
p_0 = m^4 |L|^{-3} ((b^0)^2  - \lambda^2 \|\mathbf{b}'\|^2 - 
\lambda^2 \|\mathbf{b}''\|^2) b^0,
\end{equation}
\begin{equation} \label{IP2}
\mathbf{p}' = m^4 |L|^{-3}\lambda^2 
\left(((b^0)^2 - \lambda^2 \|\mathbf{b}'\|^2 
+ \lambda^2 \|\mathbf{b}''\|^2) \mathbf{b}' 
- 2 \lambda^2 (\mathbf{b'} \cdot \mathbf{b}'') \mathbf{b}''\right),
\end{equation}
\begin{equation} \label{IP3}
\mathbf{p}'' = m^4 |L|^{-3} \lambda^2
\left(((b^0)^2 + \lambda^2 \|\mathbf{b}'\|^2 
- \lambda^2 \|\mathbf{b}''\|^2) \mathbf{b}'' 
- 2 \lambda^2 (\mathbf{b''} \cdot \mathbf{b}') \mathbf{b}' \right).
\end{equation}
The quantities $b^0$, $\mathbf{b}'$ and $\mathbf{b}''$ defined in this way satisfy a scalar primary constraint, that, since the powerful symmetry under $GL(4, \mathbf{R})$ has been broken, is rather difficult to calculate explicitly. 

The equations (\ref{IP1}, \ref{IP2}, \ref{IP3}) can be obtained from eqs.\  (\ref{P1}, \ref{P3}, \ref{P4}) by means of the substitution $\mathbf{b} = 0$. By performing the same substitution, however, eq.\ (\ref{P2}) does not give $\mathbf{p} = 0$ and eq.\ (\ref{Prim2}) does not give the scalar constraint of the present model. The quantities $p^0$, $\mathbf{p}'$ and $\mathbf{p}''$ satisfy eq.\ (\ref{Prim2}) together with the (wrong) value of $\mathbf{p}$ given by eq.\ \ (\ref{P2}). This remark is sufficient to assure the validity of the inequalities (\ref{Inequal}).

A different model of the kind b) for spinning particles can be found by means of the presymplectic formalism starting from the constraints (\ref{Prim2}) and $\mathbf{p} = 0$. Taking the second constraint into account, the scalar constraint takes the simpler form 
\begin{equation} \label{SConstr}
\Phi_0 =  \left((p_0)^2  - \lambda^{-2} \|\mathbf{p}'\|^2
- \lambda^{-2} \|\mathbf{p}''\|^2 \right)^2 
- 4 \lambda^{-4} \|\mathbf{p}' \times \mathbf{p}''\|^2 - m^4 = 0.
\end{equation}
As we have already observed, this is not the scalar constraint of the preceding model. It is evident that, in this case too, the inequalities (\ref{Inequal}), which are consequences of eq.\ (\ref{Prim2}), are satisfied.

From eq.\ (\ref{Dyn5}) we obtain
\begin{equation} \label{Inverse5}
b^0 = \alpha \left((p_0)^2 - \lambda^{-2} \|\mathbf{p}'\|^2 
- \lambda^{-2} \|\mathbf{p}''\|^2 \right) p_0,
\end{equation}
\begin{equation} \label{Inverse6}
\mathbf{b}' = 
\alpha \lambda^{-2} \left(((p_0)^2 - \lambda^{-2} \|\mathbf{p}'\|^2  
+ \lambda^{-2} \|\mathbf{p}''\|^2) \mathbf{p}' 
- 2 \lambda^{-2} (\mathbf{p}' \cdot \mathbf{p}'') \mathbf{p}'' \right),
\end{equation}
\begin{equation} \label{Inverse7}
\mathbf{b}'' =
\alpha \lambda^{-2} \left(((p_0)^2 + \lambda^{-2} \|\mathbf{p}'\|^2 
- \lambda^{-2} \|\mathbf{p}''\|^2) \mathbf{p}'' 
- 2\lambda^{-2} (\mathbf{p}'' \cdot \mathbf{p}') \mathbf{p}' \right).
\end{equation}
The vector $\mathbf{b}$ and the coefficient $\alpha$ remain undetermined.

We remark that eqs.\  (\ref{Inverse5}, \ref{Inverse6}, \ref{Inverse7}) coincide with eqs.\ (\ref{Inverse1}, \ref{Inverse3}, \ref{Inverse4}) for $\mathbf{p} = 0$, while eq.\ (\ref{Inverse2}) gives a specific value of $\mathbf{b}$, which in the present model is not necessarily correct. It follows, however, that the quantities $b^0, \mathbf{b}', \mathbf{b}''$ are the components of a vector in the seven-dimensional subspace $\mathcal{T}_7$ of $\mathcal{T}$, which belongs to a convex cone that is the projection of $\mathcal{T}^+$ on $\mathcal{T}_7$. In particular we get the the second and the third inequalities of eq.\ (\ref{Bounds}), which give separate bounds for the pseudo-acceleration and the angular velocity. The joint upper bound for these quantities is not given by eq.\ (\ref{Bound}) and its calculation is rather complicated.

The two models introduced above are different in their details, but have some common features that they share with a wider class of similar models. From the vector primary constaint and the dynamical equation (\ref{PDot2}) we obtain
\begin{equation} \label{Secondary}
p_0 \mathbf{b}'' = b^0 \mathbf{f}. 
\end{equation}
According to the general results of section 3.2, the scalar primary constraint is conserved and does not generate other secondary constraints. Eq.\ (\ref{PDot3}) implies that $\mathbf{p}'$ is constant, as it also follows from the Noether theorem. Eq.\ (\ref{PDot1}) is a consequence of the other equations and of the scalar constraint. Eq.\ (\ref{PDot4}) can be used to determine the quantity $\mathbf{b}$.

If there are no external forces, namely $\mathbf{f}  = 0$, we have $\mathbf{b}'' = 0$. In the first model, eq.\ (\ref{IP3}) gives the vector secondary constraint $\mathbf{p}'' = 0$. This result depends only on the symmetry of eq.\ (\ref{IP3}) under rotations and spatial inversion, since one cannot write a nonvanishing polar vector $\mathbf{p}''$ in terms of scalars and the axial vector $\mathbf{b}'$ alone. It follows that the same conclusion holds for the second model, which has the same symmetry. Then from eq.\ (\ref{PDot4}) we get $\mathbf{b} = 0$. In both the models, we easily obtain
\begin{equation} 
(p_0)^2  = m^2 + \lambda^{-2} \|\mathbf{p}'\|^2.
\end{equation}
Since in a zero-momentum frame, $p_0$ is the mass and $\|\mathbf{p}'\| = \sigma$ is the spin of the particle, this equation gives a relation between mass and spin.

If the maximal acceleration is an effect of quantum gravity,  $\lambda$ is of the order of Planck's length. If, as required by quantum theory for a spinning particle, $\sigma$ is of the order of $\hbar$, the particle must have a mass of the order of Planck's mass and it cannot be observed. In order to describe the observed low mass spinning particles, one has to introduce a different model, as in the subsection 7.5.  

We have seen that if $\mathbf{f}  = 0$ there are four primary constraints, three secondary constraints and one gauge variable. It follows that the phase space has dimension twelve. If there are external fields and $\hat \mathbf{B} \neq 0$, $\mathbf{f}$ depends on $\mathbf{b}$ and if we use eq.\ (\ref{PDot4}) to eliminate this variable, we obtain an equation that contains the derivatives $\dot \mathbf{p}''$ and cannot be considered as a vector secondary constraint. However, we can deduce the equation
\begin{equation} 
p_0 \hat \mathbf{B} \cdot \mathbf{b}'' = b^0 \hat \mathbf{B} \cdot \hat \mathbf{E},
\end{equation}
and by means of eqs.\ (\ref{Inverse5}) and  (\ref{Inverse7}) (or of analogous equations of the other model) we find a true secondary constraint. With respect to the free case, the number of constraints has decreased by two units and the dimension of the phase space is fourteen. The points where $\hat \mathbf{B} = 0$ are singular points of the evolution space. The nature of the additional ``internal''degree of freedom is clarified in the next subsection, where a simpler model is considered.

The quantity $(b^0)^{-1 }\mathbf{b}'$ is the angular velocity of the precession of the frame $s(\tau)$ with respect to a Fermi-Walker transported frame. If the conserved quantity $\mathbf{p}'$ vanishas, in the second model from eq.\ (\ref{Inverse6}) we have $\mathbf{b}' = 0$. This result too depends only on the symmetry of eq.\ (\ref{Inverse6}) under rotations and spatial inversion and it is true also in the first model. This means that, if the spin is very small and has no influence on the dynamics, the frame $s(\tau)$ and the spin are Fermi-Walker transported. In these conditions one does not find the correction to the Thomas precession formula which has been proposed, on the basis of a different model, in ref.\ \cite{Schuller1}. 

If the spin is large, we see from eq.\ (\ref{Inverse6}) that it has a precession around the direction of $\mathbf{p}''$, a behaviour that reminds the motion of a symmetric top. The precession due to the magnetic moment is not considered in this model.

\subsection{A model for a spinless particle.}

Now we consider a simple model of the kind b) based on the Lagrangian
\begin{equation} 
L = - m \left((b^0)^2 - \lambda^2 \|\mathbf{b}''\|^2 \right)^{1/2},
\end{equation}
that can be obtained from the Lagrangian (\ref{Lagr8}) by eliminating the terms that contain $\mathbf{b}'$.  

This can be considered as an independent model or as an approximation of the models described in the preceding subsection. We have remarked that, in a quantized model,  the excitation of the spin degrees of freedom requires a large amount of energy and if the available energy is much smaller than $\hbar \lambda^{-1}$, these degrees of freedom are ``frozen'', in the same way as, for instance, the nuclear degrees of freedom are frozen in low energy atomic physics.

We obtain the vector constraints
\begin{equation}  \label{VConstr}
\mathbf{p} = 0, \qquad \mathbf{p}' = 0,
\end{equation} 
namely, $s(\tau)$ is a zero-momentum frame and the particle is spinless. The other momenta are given by
\begin{eqnarray} \label{PP}
&p_0 = m b^0 \left((b^0)^2 - \lambda^2 \|\mathbf{b}''\|^2 \right)^{-1/2},& \nonumber \\
&\mathbf{p}'' = m \lambda^2 \left((b^0)^2 - \lambda^2 \|\mathbf{b}''\|^2 
\right)^{-1/2} \mathbf{b}''&
\end{eqnarray}
and satisfy scalar primary constraint
\begin{equation}  \label{Constr2}
(p_0)^2 - \lambda^{-2} \|\mathbf{p}''\|^2 = m^2,
\end{equation}
which is conserved and does not give rise to secondary constraints. We consider only solutions with $p_0 > 0$ and $b^0 > 0$. From the preceding formulas one obtains the relations
\begin{equation} \label{BB}
\mathbf{a} = (b^0)^{-1} \mathbf{b}'' = \lambda^{-2} (p_0)^{-1} \mathbf{p}'' =
- \lambda^{-2} \mathbf{y} =
\lambda^{-1} (\|\mathbf{p}''\|^2 + \lambda^{2} m^2)^{-1/2} \mathbf{p}'',
\end{equation}
which show that the pseudo-acceleration $\|\mathbf{a}\|$ has the upper bound $\lambda^{-1}$ and the distance $\|\mathbf{y}\|$ of the center of mass from the origin has the upper bound $\lambda$.  

The same results can be obtained in the presymplectic formalism from eq.\ (\ref{Dyn5}) starting from the constraints (\ref{VConstr}) and (\ref{Constr2}). The scalar constraint (\ref{Constr2}) can also be obtained by substituting $\mathbf{p}' = 0$ into the constraint (\ref{SConstr}) and this means that the two models discussed in the preceding subsection coincide in the zero-spin approximation.

The relation
\begin{equation} 
p_0 = m \left(1 - \lambda^{-2} \|\mathbf{y}\|^2 \right)^{-1/2}
\end{equation}
shows that $p_0$ can be interpreted as confining potential that binds the charge to the center of mass. When a force is applied to the charge, $\|\mathbf{y}\|$ and $p_0$ increase and this explains intuitively why the pseudo-acceleration remains bounded.

Eq.\ (\ref{PDot1}) follows from the constraint (\ref{Constr2}) and the other equations. From eq.\ (\ref{PDot2}) we obtain the eq.\ (\ref{Secondary}) and eq.\ (\ref{PDot3}) is trivially satisfied. Eq.\ (\ref{PDot4}) takes the form
\begin{equation} \label{DynE1}
\dot \mathbf{p}'' = p_0 \mathbf{b} - \mathbf{b}' \times \mathbf{p}''
\end{equation}
and can be used to determine the vector $\mathbf{b}$.

The time evolution of the quantities $b^0$ and $\mathbf{b}'$ is not determined by the dynamical equations and the model has gauge invariances. If we like, we can impose the gauge fixing conditions $b^0 = 1$ (a particular choice of the parameter $\tau$) and $\mathbf{b}' = 0$ (the frame is Fermi-Walker transported) and we get the simplified equations
\begin{equation} \label{DynE2}
\mathbf{p}''  = \lambda^2 \mathbf{f}, \qquad
\dot \mathbf{p}'' = p_0 \mathbf{b}.
\end{equation}

If there is a gravitational field, but no other external fields, from eqs.\ (\ref{Dyn6}), (\ref{EB}) and (\ref{EBf}), we see that $\mathbf{p}' = \mathbf{p}'' = 0$ implies $\mathbf{f} = 0$. It follows that, if initially $\mathbf{p}'' = 0$, we have $\mathbf{f} = \mathbf{p}'' = \mathbf{b}'' = \mathbf{b} = 0$ for all the values of $\tau$ and $s(\tau)$ is a parallel transported rest frame. The solution describes an ordinary spinless particle with mass $p_0 = m$ moving according to the laws of general relativity, but in general there are other different solutions with $\mathbf{p}'' \neq 0$.  If there is no external field, we have necessarily $\mathbf{f} = \mathbf{p}'' = 0$ and there are no other solutions. In this case, the phase space has dimension six.

In general, as in the preceding subsection, we have to remember that the force $\mathbf{f}$ may depend on the velocity $\mathbf{b}$. By eliminating $\mathbf{b}$ from the equations (\ref{DynE2}) we obtain 
\begin{equation} 
\mathbf{p}''  = \lambda^2 (\hat \mathbf{E} 
+ (p_0)^{-1} \dot \mathbf{p}'' \times \hat\mathbf{B}).
\end{equation}
We see that if $\hat \mathbf{B} \neq 0$, the component of $\mathbf{p}''$ parallel to this vector is determined, while the two normal components can be chosen arbitrarily in the initial conditions and the equation determines their time evolution. We have seven primary constraints, one secondary constraint and four arbitrary gauge variables. It follows that the dimension of the phase space is eight. If $\hat \mathbf{B} = 0$, all the three components of $\mathbf{p}''$ are determined and the phase space has dimension six, as in the free case.

This model too has some problems. If $\mathbf{f}$ is small but oscillates very rapidly,  $p_0$ remains near to $m$ but the velocity $\mathbf{b}$ of the charge can be larger than $c = 1$. If, instead of considering a test particle, we take into account terms proportional to $e^2$, we have to consider a large radiated energy due to the acceleration and also Cerenkov radiation if the velocity is larger than $c$. 

Now we consider, as an exercise, a charged particle in a constant magnetic field in the absence of gravitation. We assume that in the fixed frame $\hat s$ the only nonvanishing component of the electromagnetic field is $F_{12} = - B^3 = \|\mathbf{B}\| = B$ and we try to find a particular kind of periodic solutions of the form 
\begin{equation} \label{Circum}
s(\tau) = g(\tau) \hat s = \exp(\zeta A_{[20]}) \exp(r A_1) 
\exp(\omega\tau A_{[12]} + \tau A_0) \hat s,
\end{equation}
which represents a frame with its origin moving with angular velocity $\omega$ along a circumference of radius $r$ and with the $x^1$-axis parallel to the radius. Finally, the frame is boosted along the $x^2$-axis with a rapidity $\zeta$. Note that the  gauge fixing is different from the one suggested above. The quantities $b^{\alpha}$ defined by eq.\ (\ref{Deriv}) can be found by computing the derivative $\dot g g^{-1}$. One finds in this way
\begin{eqnarray} \label{BBB}
&b^{0} =  \gamma (1 - \omega r \beta), \qquad 
\mathbf{b} = \gamma (\omega r - \beta) (0, 1, 0),& \nonumber \\
&\mathbf{b}' = (0, 0, \gamma \omega), \qquad
\mathbf{b}'' = (- \gamma \beta \omega, 0, 0),&
\end{eqnarray}
where
\begin{equation} 
\beta = \tanh \zeta, \qquad \gamma = \cosh \zeta.
\end{equation}

After the final Lorentz boost, the nonvanishing components of the electromagnetic field take the form
\begin{equation} 
\mathbf{E} = B \gamma \beta (- 1, 0, 0), \qquad 
\mathbf{B} = B \gamma (0, 0, -1),
\end{equation}
and we have
\begin{equation} 
b^0 \mathbf{f} = (- eB \omega r, 0, 0).
\end{equation}
We also obtain
\begin{equation} \label{PP1}
p_0 = m \left(1 - \lambda^2 \omega^2 \beta^2 
(1 - \omega r \beta)^{-2} \right)^{-1/2},
\end{equation}
\begin{equation} \label{COM1}
\mathbf{y} = - (p_0)^{-1} \mathbf{p}'' =  
\lambda^2 \omega \beta (1 - \omega r \beta)^{-1} (1, 0, 0).
\end{equation}

From eqs.\ (\ref{Secondary}) and (\ref{DynE1}) we obtain the conditions
\begin{equation} \label{Equa1}
(\omega r - \beta)(\omega r \beta -1) = \lambda^2 \omega^2 \beta,
\end{equation}
\begin{equation} \label{Equa2}
p_0 \gamma \beta = e B r.
\end{equation}
The nonrelativistic approximation, in which  $\omega r$ and $\beta$ are small, gives
\begin{equation} \label{NRel}
eB = m \omega (1 - \lambda^2 \omega^2)^{-1}.
\end{equation}
This expression, valid for arbitrary $\lambda$, gives a correction to the cyclotron frequency.

From eqs.\ (\ref{COM1}) and (\ref{Equa1}) we have
\begin{equation} 
\mathbf{y} = (\beta \omega ^{-1} - r) (1, 0, 0)
\end{equation}
and from eq.\ (\ref{BB}) we obtain the inequality
\begin{equation} 
|\beta \omega ^{-1} - r| < \lambda,
\end{equation}
which also assures that $p_0$ is real. We see that the center of mass moves with velocity $|\beta| < 1$ on a circumference of radius $|\beta \omega^{-1}|$. If $\beta \omega^{-1} < 0$, the center of mass and the charge have opposite positions with respect to the common centre of rotation and their distances from the center are smaller than $\lambda$.

We assume that $eB > 0$ and $r > 0$ and it follows that that $0 < \beta < 1$. For fixed values of $\lambda$ and $\beta$, the inequalities given above, together with eq.\ (\ref{Equa1}) define a set in the space of the variables $\omega$ and $r$ that is the union of two connected components. In the two components the various quantities vary in the intervals defined by the following two sets if inequalities:
\begin{equation} \label{Set1}
0 < \omega < \lambda^{-1}, \qquad \infty > r > 0, \qquad
\beta > \omega r >0,
\end{equation}
\begin{equation} \label{Set2}
- \infty < \omega < - \lambda^{-1}, \qquad \lambda > r > 0, \qquad
- \infty < \omega r <0.
\end{equation}

The solutions of the first set approach the usual classical solutions when $\omega r \approx \beta $, and $\lambda \omega$ is small. Disregarding higher powers of the last quantity,  we obtain 
\begin{eqnarray} \label{Approx}
&\omega r \approx \beta (1 - \lambda^2 \omega_0^2 \gamma^2), \qquad
p_0 \approx m (1 + 2^{-1} \lambda^2 \omega_0^2 \gamma^4 \beta^2),& 
\nonumber \\
&\omega \approx \omega_0 \left(1 -  \lambda^2 \omega_0^2 \gamma^2
(1 + 2^{-1} \gamma^2 \beta^2) \right), \qquad 
\omega_0 = e B (m \gamma)^{-1},&
\end{eqnarray}
namely small corrections, proportional to $(\lambda \omega_0)^2$, to the classical formulas that describe a point charged particle in a magnetic field. The ``internal'' degrees of freedom do not appear in this approximation.

The solutions of the second set have no classical analog. In particular, $\omega$ takes negative values, namely the angular velocity vector is parallel to $e \mathbf{B}$, while in the classical case it is antiparallel.  We have seen that in this case both the center of mass and the charge have distances smaller than $\lambda$ from the center of rotation. For some of these solutions we have $|\omega|r > 1$, namely the charge rotates faster than light. These solutions involve essentially the new ``internal'' degrees of freedom that, as we have observed above, appear in the model when a magnetic field is present.

We see from eq.\ (\ref{Set2}) that $|\omega|$ is very large and it would be difficult to observe these motions in experiments, specially in a quantized version of the model, in which very high energy quanta have to be exchanged. One may say that
for energies much smaller than $\hbar \lambda^{-1}$ the internal degrees of freedom are ``frozen'', as the spin degrees of freedom of the preceding subsection.

There are also solutions which involve both the classical ``external'' degrees of freedom and the new ``internal'' ones. They are presumably multiply periodic and their treatment requires more refined methods of analytic mechanics. 

\subsection{A model with a magnetic dipole moment.}

In the present section we discuss a model that describes a particle with spin, magnetic dipole and maximal pseudo-acceleration.  We start from the presymplectic model described in the section 5.4 and, in  order to introduce a maximal pseudo-acceleration, we modify the constraints in such a way that for $\boldsymbol{\sigma} = \boldsymbol{\mu} = 0$ we obtain eq.\ (\ref{VConstr}) and (\ref{Constr2}) of the subsection 7.3, namely we put 
\begin{equation} 
\Phi_0 = p_0 - (m^2 + \lambda^{-2} \|\mathbf{p}''\|^2)^{1/2} + \boldsymbol{\mu} \cdot \mathbf{B}, \qquad \mathbf{p}' = \boldsymbol{\sigma},   
\qquad \mathbf{p} = 0.
\end{equation}
From eq.\ (\ref{Dyn5}) we obtain
\begin{equation} 
b^0 = \alpha^0, \qquad
\mathbf{b}'' = b^0 \lambda^{-2} (m^2 + \lambda^{-2} \|\mathbf{p}''\|^2)^{-1/2} \mathbf{p}'',
\end{equation}
while the vectors $\mathbf{b}$ and $\mathbf{b}'$ remain undetermined. 

As in the model of section 5.4, the last term in eq.\ (\ref{Dyn8}) has to be taken into account and we obtain the dynamical equations (\ref{CPDot1}--\ref{CPDot3}), while eq.\ (\ref{CPDot4}) has to be replaced by
\begin{equation} \label{CPDot6}
\dot \mathbf{p}'' = p_0 \mathbf{b} 
- \mathbf{b}' \times \mathbf{p}'' + \mathbf{b}'' \times \boldsymbol{\sigma}
+ b^0 \boldsymbol{\mu} \times \mathbf{E}.
\end{equation}
The discussion of these equations is similar to the one given in sections 5.4 and 7.3. Eq.\ (\ref{CPDot1}) is a consequence of the scalar constraint, eq.\ (\ref{CPDot2}) determines the pseudo-acceleration $(b^0)^{-1} \mathbf{b}''$ and eq.\ (\ref{CPDot6}) determines the vector $\mathbf{b}$ and can be used to eliminate it from the other equations. By fixing the gauge related to the rotations of the frame around $\boldsymbol{\sigma}$, we can replace eq.\ (\ref{CPDot3}) by the stronger equation (\ref{CPDot5}). 

If $\hat \mathbf{B} \neq 0$, only the projection of eq.\ (\ref{CPDot2}) on its direction gives rise to a secondary constraint, otherwise all the three components of this equation generate secondary constraints. Since we have seven primary constraints and two arbitrary gauge variables, in the first case the dimension of the phase space is ten, while in the second case it is eight, as it is expected for an ordinary spinning particle \cite{Souriau}. Note that the second singularity described by eq.\ (\ref{Sing}) has disappeared as a consequence of the introduction of the maximal acceleration.

For this model too, we consider a particle in a constant magnetic field. Eq.\ (\ref{Circum}) is not sufficiently general and we look for a solution of the kind
\begin{eqnarray} \label{Circum2}
&s(\tau) = g(\tau) \hat s& \nonumber \\
&= \exp(\eta A_3) \exp(\rho \tau A_{[12]}) \exp(\zeta A_{[20]}) \exp(r A_1) \exp(\omega\tau A_{[12]} + \tau A_0) \hat s, \quad&
\end{eqnarray}
where $\rho, \zeta, \omega$ are constants and $\eta$ is a function of the time parameter $\tau$. By proceeding as in section 7.3 we find
\begin{eqnarray} \label{BBB2}
&b^{0} =  \gamma (1 - \omega r \beta),& \nonumber \\
&\mathbf{b} = \gamma (\omega r - \beta)
(\sin \rho \tau, \cos \rho \tau, 0) + (0, 0, \dot \eta),& \nonumber \\
&\mathbf{b}' = (0, 0, \rho + \omega \gamma), \qquad
\mathbf{b}'' = -\omega \gamma \beta (\cos \rho \tau, -\sin \rho \tau, 0).&
\end{eqnarray}

In the frame $s(\tau)$ the electromagnetic field takes the form
\begin{equation} 
\mathbf{E} = B \gamma \beta (- \cos \rho \tau, \sin \rho \tau, 0), \qquad 
\mathbf{B} = B \gamma (0, 0, -1),
\end{equation}
and it follows
\begin{equation} 
b^0 \mathbf{f} = eB \omega r (-\cos \rho \tau, \sin \rho \tau, 0).
\end{equation}
We also obtain
\begin{equation} \label{Mag0}
p_0 = m \left(1 - \lambda^2 \omega^2 \beta^2 
(1 - \omega r \beta)^{-2} \right)^{-1/2} + \mu^3 B \gamma.
\end{equation}

From the dynamical equations (\ref{CPDot2}), (\ref{CPDot5}) and (\ref{CPDot6}) we obtain eq.\ (\ref{Equa2}) and the relations
\begin{equation} \label{Mag1}
\rho + \omega \gamma = g e (2m)^{-1} B \gamma^2 (1 - \omega r \beta),
\end{equation} 
\begin{eqnarray} \label{Mag2}
&p_0 \left( \lambda^2 \omega^2 \beta - 
(\beta - \omega r) (1 - \omega r \beta) \right)&
\nonumber \\
&= \mu^3 B \gamma \beta \left(\lambda^2 \omega^2 + (1 - \omega r \beta)^2 \right)
- \sigma^3 \omega \beta (1 - \omega r \beta)&,
\end{eqnarray} 
\begin{equation} 
p_0 \dot \eta = \gamma \beta \left( \omega - ge(2m)^{-1}B \gamma (1 - \omega r \beta) \right) (\sigma^1 \sin \rho \tau + \sigma^2 \cos \rho \tau).
\end{equation} 
The last equation describes an oscillation in the direction of the magnetic field.

As we see from eqs.\ (\ref{Circum2}) and (\ref{BBB2}), the quantity $\rho + \omega \gamma$ determined by eq.\ (\ref{Mag1}) is the angular velocity of the moving frame with respect to a Fermi-Walker transported frame. The quantity $\rho + \omega$ is the angular velocity of the moving frame with respect to the fixed frame. Their difference $\omega (1 - \gamma)$ is the angular velocity of the Thomas precession. The momentum of the particle with respect to a fixed frame rotates with angular velocity $\omega$ and $\rho$ is the difference between the angular velocities of the moving frame (namely of the spin) and the momentum.

If $\sigma^3 = \mu^3 = 0$, namely if the magnetic moment is perpendicular to the magnetic field, eqs.\ (\ref{Mag0}) and (\ref{Mag2}) coincide to eqs.\ (\ref{PP1}) and (\ref{Equa1}) of section 7.3 and the evolution of the variables $\omega, \beta, r$ is not affected by spin and magnetic moment. If we use the approximation (\ref{Approx}),  we obtain
\begin{equation}
\rho + \omega \gamma \approx 
2^{-1} g \omega_0 \gamma (1 + \lambda^2 \omega_0^2 \gamma^4 \beta^2),
\end{equation} 
and
\begin{equation}
\rho = 2^{-1} (g - 2) \omega_0 \gamma + O(\lambda^2 \omega_0^2).
\end{equation} 
The last formula shows that, for $\lambda = 0$, $\rho$ is proportional to $g - 2$, a remark which is exploited for precise experimental measurements of the anomalous magnetic moment of Dirac particles. In the experimental situations the maximal acceleration corrections are completely negligible, even if we put $\lambda = \hbar m^{-1}$ (the Compton wave length).

\subsection{The model of section 2 revisited.}

We consider only one model of the kind a), which is just a reformulation of the model treated in section 2, similar to the one given in ref.\ \cite{NFS}. We introduce the Lagrange multipliers $\boldsymbol{\eta}$ and we start from the Lagrangian
\begin{equation} \label{Lagr6}
L = - m \left((b^0)^2 - \lambda^2 \|\mathbf{b}''\|^2 \right)^{1/2} 
+ \boldsymbol{\eta} \cdot \mathbf{b}.
\end{equation} 

We obtain the equations
\begin{equation} 
\mathbf{b} = 0, \qquad \mathbf{p} = \boldsymbol{\eta}, \qquad \mathbf{p}' = 0.
\end{equation} 
The first formula means that we are dealing with rest frames and the second shows that $\mathbf{p}$ can be chosen freely in the initial conditions, even in contrast with the usual requirements on the energy-momentum spectrum. The third formula does not imply that the spin vanishes, because $\mathbf{p}'$ is not the angular momentum in a zero-momentum frame. The equations (\ref{PP}), (\ref{Constr2}) and (\ref{BB}) of section 7.3 are still valid. The last equation, however, gives an upper bound to the true acceleration $\mathbf{a}$. The same results are obtained in the presymplectic formalism starting from the constraints (\ref{Constr2}) and $\mathbf{p}' = 0$.

The scalar constraint (\ref{Constr2}) is conserved and the dinamical equation (\ref{PDot1}) follows from it. Eq.\ (\ref{PDot3}) is automatically satisfied and from eqs. (\ref{PDot2}) and (\ref{PDot4}) we obtain
\begin{equation} \label{PDot5}
\dot{\mathbf{p}} = - p_0 \mathbf{b}'' + b^0 \mathbf{f} 
- \mathbf{b}' \times \mathbf{p}, \qquad
\dot{\mathbf{p}}'' = - b^0 \mathbf{p}
- \mathbf{b}' \times \mathbf{p}''.
\end{equation}
There are no secondary constraints. 

Also in this case we have gauge symmetries. We can impose the gauge fixing conditions $b^0 = 1$ and $\mathbf{b}' = 0$ and we obtain the simpler dynamical equation
\begin{equation} \label{Exp}
\ddot{\mathbf{p}}'' = \lambda^{-2} \mathbf{p}'' - \mathbf{f}.
\end{equation}
There are four primary constraints and four arbitrary gauge variables and the phase space has dimension twelve, as we have already seen in section 2.

Since in the rest frame $s(\tau)$ we have (see section 5.1)
\begin{equation} 
\dot x^0 =1, \qquad \dot{\mathbf{x}} = 0, \qquad
\ddot{\mathbf{x}} = \mathbf{b}'',
\end{equation} 
we see that the first term in the Lagrangian (\ref{Lagr6}) coincides with the expression (\ref{Lagr3}) (see also eq.\ (\ref{Rest})). One can show that the equations given above in the frames $s(\tau)$ are equivalent to the equations given in section 2 in a fixed frame. 

It is shown in ref.\ \cite{NFS} that, in the absence of external fields, the dynamical equations can be solved by quadratures. In fact, for $\mathbf{f} = 0$ eq.\ (\ref{Exp}) has exponential solutions. For generic initial conditions, as we see from eq.\ (\ref{BB}), the acceleration and the distance $\|\mathbf{y}\|$ approach their maximum values when time increases. In a fixed frame, the center of mass moves with a constant velocity and its distance from the charge increases exponentially. In the moving frame, however, this distance remains bounded due to the Lorentz contraction. Since this situation can be avoided only for special initial conditions, the system is unstable.

As we have remarked in section 7.3, when the particle is subject to gravitational fields only, if $\mathbf{p}' = \mathbf{p}'' = 0$ we have $\mathbf{f} = 0$. It follows that $\mathbf{p}'' = \mathbf{b}'' = \mathbf{p} = 0$ is a solution which describes an ordinary spinless particle with mass $p_0 = m$ moving according to the laws of general relativity. We have seen that, even in the absence of external fields, there are other kinds of solutions and, in fact, the dimension of the phase space is larger than six.

As in the preceding sections, we consider a charged particle in a constant magnetic field and we look for periodic solutions of the kind (\ref{Circum}). The equations (\ref{BBB}--\ref{COM1}) are still valid. From the condition $\mathbf{b} = 0$ we obtain for the velocity the familiar formula
\begin{equation} \label{Beta}
\omega r = \beta < 1 
\end{equation}
and from the other equations given above we obtain
\begin{equation} 
p_0 = m \left(1 - \lambda^2 \omega^2 \gamma^4 \beta^2 \right)^{-1/2}, \qquad 
\mathbf{p} = p_0 \lambda^2 \omega^2 \gamma^4 \beta (0, 1, 0),
\end{equation}
\begin{equation} 
\mathbf{y} = - (p_0)^{-1} \mathbf{p}'' =
\lambda^2 \omega \gamma^2 \beta (1, 0, 0)
\end{equation}
\begin{equation} 
e B = m \omega \gamma
\left(1 + \lambda^2 \omega^2 \gamma^4 \right)
\left(1 - \lambda^2 \omega^2 \gamma^4 \beta^2 \right)^{-1/2}.
\end{equation}
We see that, for $eB > 0$, negative values of $\omega$ are not admitted and that the center of mass rotates on a circunference with radius larger than $r$. In order to have a real $p_0$ and $\|\mathbf{y}\| < \lambda$, we have to impose the condition
\begin{equation} \label{Up}
\lambda \omega < \gamma^{-2} \beta^{-1}.
\end{equation}
For $\lambda$ and $\beta$ fixed, the relations between $B$, $\omega$ and $p_0$ are monotonic and when $\omega$ approaches its upper bound (\ref{Up}), $B$ and $p_0$ tend to infinity. One can see easily that in this limit the energy-momentum becomes spacelike.
 
Disregarding higher powers of $\lambda \omega$, we obtain the formula
\begin{equation} \label{Approx2}
\omega \approx \omega_0 \left(1 - \lambda^2 \omega_0^2 \gamma^4
(1 + 2^{-1} \beta^2) \right),
\end{equation}
that coincides with eq.\ (\ref{Approx}) in the nonerelativistic limit $\gamma \to 1$.  For arbitrary $\lambda$ the nonrelativistic approximation gives
\begin{equation} 
eB = m \omega (1 + \lambda^2 \omega^2).
\end{equation}
which is different from eq. (\ref{NRel}).

The additional degrees of freedom, which are present in this model, do not appear in the periodic solutions because, as we have seen, they have an exponential character.

\end{document}